\documentclass[default,iicol]{sn-jnl}% Default with double column layout

\usepackage{xcolor}
\usepackage{comment}
\usepackage{amsmath}
\usepackage{amssymb}
\usepackage{tikz}

\usepackage{hyperref}

\usepackage{program}%

\usepackage{subfigure}
\usepackage{caption}

\usepackage{tikz,pgfplotstable,filecontents}

\pgfplotstableread{A.dat}\A
\def\nrows{29}
\def\ncols{29}

\pgfplotstableread{A_chol.dat}\cholA
\def\nrows{29}
\def\ncols{29}

\pgfplotstableread{A_perm.dat}\permA
\def\nrows{29}
\def\ncols{29}

\pgfplotstableread{A_perm_chol.dat}\permCholA
\def\nrows{29}
\def\ncols{29}

\definecolor{myblue}{HTML}{3F51B5} %Blue
\definecolor{mylightblue}{HTML}{9FA8DA}

\definecolor{myblue}{RGB}{30,125,186}
\definecolor{myred}{RGB}{192, 50, 33}

\usetikzlibrary{shapes.misc}
\tikzset{cross1/.style={cross out, draw=black, thick, minimum size=2*(#1-\pgflinewidth), inner sep=0pt, outer sep=0pt},
        cross1/.default={2.5pt}}

\tikzset{cross2/.style={cross out, draw=gray!100, thick, minimum size=2*(#1-\pgflinewidth), inner sep=0pt, outer sep=0pt},
        cross2/.default={2.5pt}}

\tikzset{cross3/.style={cross out, draw=red, ultra thick, minimum size=2*(#1-\pgflinewidth), inner sep=0pt, outer sep=0pt},
        cross3/.default={4pt}}

\renewcommand{\vec}[1]{\boldsymbol{#1}}
\renewcommand{\th}{\boldsymbol{\theta}}
\renewcommand{\x}{\boldsymbol{x}}
\renewcommand{\y}{\boldsymbol{y}}
\renewcommand{\Q}{\boldsymbol{Q}}
\renewcommand{\L}{\boldsymbol{L}}

\jyear{2021}%

\theoremstyle{thmstyleone}%
\theoremstyle{thmstyletwo}%

\theoremstyle{thmstylethree}%

\begin{document}

% TODO: find better title!! \title[Article Title]{Scalable
% high-performing approximate Bayesian inference using parallelized
% integrated nested Laplace approximations}
\title[Article Title]{Parallelized integrated nested Laplace
    approximations for fast Bayesian inference}

%% =============================================================%%
%% Prefix -> \pfx{Dr} GivenName -> \fnm{Joergen W.} Particle ->
%% \spfx{van der} -> surname prefix FamilyName -> \sur{Ploeg} Suffix
%% -> \sfx{IV} NatureName -> \tanm{Poet Laureate} -> Title after name
%% Degrees -> \dgr{MSc, PhD} \author*[1,2]{\pfx{Dr} \fnm{Joergen W.}
%% \spfx{van der} \sur{Ploeg} \sfx{IV} \tanm{Poet Laureate} \dgr{MSc,
%% PhD}}\email{iauthor@gmail.com}
%% =============================================================%%

% lisa, janet, olaf, haavard
\author*[1]{\fnm{Lisa}
    \sur{Gaedke-Merzh\"{a}user}}\email{lisa.gaedke.merzhaeuser@usi.ch}

\author[2]{\fnm{Janet} \sur{van
        Niekerk}}\email{janet.vanniekerk@kaust.edu.sa}

\author[1]{\fnm{Olaf} \sur{Schenk}}\email{olaf.schenk@usi.ch}

\author[2]{\fnm{H\r{a}vard} \sur{Rue}}\email{haavard.rue@kaust.edu.sa}

\affil*[1]{\orgdiv{Faculty of Informatics}, \orgname{Universita della
        Svizzera italiana}, \orgaddress{\city{Lugano},
        \postcode{6900}, \country{Switzerland}}}

\affil[2]{\orgdiv{CEMSE Division}, \orgname{King Abdullah University
        of Science and Technology},\orgaddress{\city{Thuwal},
        \postcode{23955-6900}, \country{Saudi Arabia}}}

% affil[3]{\orgdiv{Department}, \orgname{Organization},
% \orgaddress{\street{Street}, \city{City}, \postcode{610101},
% \state{State}, \country{Country}}}


\abstract{There is a growing demand for performing larger-scale
    Bayesian inference tasks, arising from greater data availability
    and higher-dimensional model parameter spaces. In this work we
    present parallelization strategies for the methodology of
    integrated nested Laplace approximations (INLA), a popular
    framework for performing approximate Bayesian inference on the
    class of Latent Gaussian models. Our approach makes use of nested
    OpenMP parallelism, a parallel line search procedure using robust
    regression in INLA’s optimization phase and the state-of-the-art
    sparse linear solver PARDISO. We leverage mutually independent
    function evaluations in the algorithm as well as advanced
    % extremely optimized
    sparse linear algebra techniques. This way we can flexibly utilize
    the power of today's multi-core architectures. We demonstrate the
    performance of our new parallelization scheme on a number of
    different real-world applications. The introduction of parallelism
    leads to speedups of a factor 10 and more for all larger models.
    % Showing an average speedup of ... compared to the previous
    % implementation and \textit{and something scalability}.
    Our work is already integrated in the current version of the
    open-source R-INLA package, making its improved performance
    conveniently available to all users.}

\keywords{Bayesian inference, INLA, Parallelism, OpenMP}

%% \pacs[JEL Classification]{D8, H51}

%% \pacs[MSC Classification]{35A01, 65L10, 65L12, 65L20, 65L70}

\maketitle

%%%%%%%%%%%% GENERAL TODOs %%%%%%%%%%%%

% TODO: potential conflicts of naming : x latent paramter space, as
% well as solve Qx = b but here its actually the right thing .. but
% still for now both x

% TODO: think about where and how little graphics could be added to
% make illustrate things

%%%%%%%%%%%%%%%%%%%%%%%%%%%%%%%%%%%%%%

\section{Introduction}\label{sec:intro}
%%% NOTE: no subheadings in introduction
%%%% concrete beginning General bla bla
% Bayesian inference is becoming an increasingly used approach for
% modeling a wide variety of problems and phenomena in fields like
% statistics, applied mathematics and machine learning. While the
% further development of fast and reliable methods for the often
% challenging task of performing Bayesian inference is an active area
% of research \textit{cite}, this is matched by a growing availability
% of data and continuously evolving statistical models. Among the
% different methods for performing Bayesian inference Markov Chain
% Monte Carlo is usually considered as the gold standard in terms of
% accuracy. Their theoretical foundations show that the comprised
% methods can reach an arbitrary level of accuracy for arbitrary
% \textit{posterior} distributions. Unfortunately the necessary
% sampling process is often very demanding or even practically
% infeasible in terms of complexity and/or computational time
% \textit{cite}. Over the years alternative techniques have emerged to
% ideally provide equally good results at shorter run time and/or less
% necessary tuning. Among which there is the framework of integrated
% nested Laplace approximations.
%%%%%
The methodology of integrated nested Laplace approximations (INLA) has
become a widely spread framework for performing complex Bayesian
inference tasks, see e.g.~\cite{rue2017bayesian,
    opitz2017latent,art643,rue2017bayesian}. INLA is applicable to a
wide subclass of additive Bayesian hierarchical models. It employs a
deterministic approximation scheme that relies on Gaussian Markov
random fileds (GMRFs) in the latent parameter space which enables the
usage of efficient sparse linear algebra
techniques~\cite{rue2009approximate}. A user-friendly implementation
of INLA is available in the form of an R-package under the same name,
referred to as R-INLA. It can be downloaded and installed as described
in \texttt{www.r-inla.org} with the complete source code available on
GitHub\footnote{\url{https://github.com/hrue/r-inla}}. Since its
inception there have been many papers exploring advancements of
theoretical concepts of the INLA methodology, also leading to a
constant evolution of its implementation. Their agglomeration forms an
impressive repertoire for fast, versatile and reliable approximate
Bayesian inference, see e.g.~\cite{lindgren2011explicit,
    martins2013bayesian}, including a wide variety of applications,
see e.g.~\cite{batomen2020vulnerable, de2019species,
    lu2018hierarchical, arisido2017bayesian, bhatt2015effect,
    konstantinoudis2021long, mielke2020, coll2019, martinez2018, isaac2020, lindenmayer2021, pimont2021, pinto2020, lillini2021, sanyal2018, shaddick2018, kontis2020}. Algorithmic concepts and improvements
concerning performance matters of the continuously growing software
library have been discussed much less, even though they are a key
component to INLA's success. In this work we want to invite the reader
to look at the INLA methodology with us from a more algorithmic point
of view.
%%%% present our work.
We will present a much more performant implementation of R-INLA making
use of OpenMP~\cite{OpenMP} parallelism and the state-of-the-art
sparse linear solver PARDISO~\cite{bollhofer2020state}.

%%%% KEY CONTRIBUTIONS
The introduction of OpenMP allows for the simultaneous execution of
multiple tasks within the algorithm operating on shared memory. We
have parallelized the computationally intensive operations within INLA
whenever the dependency structure allowed it, which also opened the
door for us to add more robust approximation strategies. Internally,
the PARDISO library is also utilizing OpenMP which leads to a double
layer of parallelism within INLA.
% In the following we first introduce the various adaptations
% throughout the approximation scheme forming the top layer
% parallelism. OpenMP function evaluations in gradient
A large quantity of the algorithm's overall runtime is spent in an
optimization routine to determine the posterior mode of the model's
hyperparameters. %using a Laplace approximation
The mode is found iteratively using a quasi-Newton method that
requires gradient information of the functional being maximized.
Approximating the gradient in each iteration is a computationally
expensive task, which we were, however, able to parallelize.
% We have parallelized this step using OpenMP which enables the
% simultaneous execution of computationally involved code sections if
% sufficient compute resources are available, subsequently
% significantly reducing the time of each gradient evaluation. OpenMP
% function evaluations with parallel line search
Additionally, INLA performs a line search procedure in every iteration
of the optimization to find a suitable parameter configuration for the
next step. Instead of sequentially looking for a good candidate we
have parallelized this process, again, using OpenMP while also
including a robust regression scheme for more stable results. This can
allow us to save time in each iteration and can also reduce the total
number of iterations required.
% OpenMP for partial inversion
To compute the posterior marginal distributions of the latent
parameters of the model partial matrix inversions are necessary. For
all computations INLA generally works with the sparse precision
matrices instead of their dense counterparts, the corresponding
covariance matrices. However, to obtain the marginal variances,
inversions of often high-dimensional precision matrices are necessary.
A time-consuming task which we have adapted to be executed in parallel
through the usage of OpenMP. The collection of these strategies forms
the first or top layer parallelism.

% PARDISO
The original implementation of INLA employs the sparse linear algebra
library TAUCS \cite{toledo2003taucs} to perform the required
numerically intensive core operations. Foremost, these include
Cholesky decompositions of matrices with recurring sparsity patterns
and the partial matrix inversions. TAUCS is a well-designed and
efficient library which operates, however, sequentially
% \footnote{\textcolor{red}{@haavard: some comment on how the parallel
% version couldn't be integrated?}}
and whose support was discontinued in the early 2000's. Additionally,
TAUCS does not include a partial inversion routine which was therefore
implemented by the authors of R-INLA, see \cite{rue2007approximate,
    rue2009approximate}, in a sequential manner. While this was done
with much thought, effort and consideration, providing reliable
results, improved parallel implementations have emerged
since~\cite{VERBOSIO201799, LI20089408}.
% \textcolor{red}{ TODO: @olaf what to cite pardiso selected
% inversion?}
They were developed by experts in the field of high-performance
computing and sparse numerical solvers. The plan of integrating
PARDISO into INLA was first described by Nierkerk et al.
in~\cite{van2019new}. The authors discuss the need for
% \textit{parallel, better, faster, newer}
the usage of faster numerical solvers within INLA to support evolving
statistical models of higher complexity and continuously growing
availability of data. They also show the efficiency and scalability
with which PARDISO performs the beforementioned numerically intensive
core operations like Cholesky decompositions and partial matrix
inversions. Especially for the latter task there are not many
performant libraries available which make PARDISO a particularly great
fit~\cite{van2019new}. Hence, the second layer of parallelism is
formed by PARDISO as well as other parallelized linear algebra
operations like matrix-matrix or matrix-vector products. We will
present performance results for three different real-world
applications, quantifying the improvement compared to previous
implementations in terms of speedup and scalability through the
various parallel OpenMP strategies. The case studies include a complex
joint survival model containing 50 hyperparameters~\cite{art698}, a
brain activation model using fMRI data with very high-dimensional
spatial domains~\cite{spencer2022spatial} and a smaller scale model
describing spatial variation in Leukemia survival
data~\cite{lindgren2011explicit}. All corresponding code and data is
publicly available.
% We have chosen three different previously published applications
% that make use of the INLA modeling framework to show case the

The rest of the paper is organized as follows. We will begin with a
brief introduction of the INLA methodology, presenting the class of
applicable models and underlying statistical concepts, with a
particular focus on the computations they entail. Additionally we will
provide an overview of the fundamental numerical methods and sparse
linear algebra operations involved in the implementation. Then we will
introduce the new parallelization schemes, before demonstrating our
improvements on the various applications.

% something something The augmented implementation has been
% continuously integrated in the R-INLA package and is fully available
% \textcolor{red}{TODO:@haavard: maybe cite version?}. While some
% features are automatically invoked others need to be set explicitly,
% for details we refer to For all case studies the source code is
% available and maybe appendix or something else? Full example code
% can be found under .

% While the applied reader might not be foremost interested in
% implementation details we still want to encourage you to read on,
% not only out of pure curiosity, but because we think that gaining a
% better understanding of the underlying implementations, will lead to
% users to a more efficient usage of INLA. All code used to obtain the
% results presented in Section () is pubicly available for
% transparency reasons but also to provide examples of how to
% efficiently use the new INLA-PARDISO features. Please see Section
% ... & Appendix (?) for details.

\section{Background}
\label{sec:background}

% \textcolor{red}{Maybe add somewhere: Everything is done using
% precision matrix. Covariance matrix only implicitly available.}

\subsection{Latent Gaussian Models}
\label{subsec:LGM}
%%% more about INLA, lets see if it ends up in introduction or
%%% background
% INLA requirements -> already mention components of hierarichal model
The INLA methodology is applicable to the class of latent Gaussian
models (LGMs). They comprise a subclass of hierarchical Bayesian
additive models among which there are many of the frequently used
statistical models, like regression, mixed or spatio-temporal models
as well as many others, see e.g.
\cite{rue2009approximate,martins2013bayesian,rue2017bayesian,art643}
for details.
% LGMs contain many of the frequently used statistical models, and
% have a particular hierarchical structure.
Each observation $y_i$ is assumed to belong to a distribution from the
exponential family and is associated with the additive linear
predictor $\eta_i$ through a link function, where $\eta_i$ is defined
as
%% covariates X_i maybe not good because of latent parameters x but
%% not sure what to replace it with ...
\begin{align}
  \eta_i = \beta_0 + \vec{\beta}^T \vec{Z}_i + \vec{u}_i (\vec{w}_i).
\end{align}
The linear predictor has fixed effects $\vec{\beta}$ with covariates
$\vec{Z}_i$ and nonlinear random effects $\vec{u}_i$ with covariates
$\vec{w}_i$. The observations are assumed to be conditionally
independent given the parameters, such that
\begin{align}
  \vec{y} \ \vert \  \vec{\eta},  \th \sim \prod_{i=1}^{m} \pi(y_i \vert \eta_i, \th),
\end{align}
where $\th$ are the hyperparamters of the model. 
The latent parameter space of the model has traditionally been defined as $\x = (\beta_0, \vec{\beta}, \vec{u}, \vec{\eta})$, but can now also be formulated without the linear predictor as $\x = (\beta_0, \vec{\beta}, \vec{u})$, which was recently presented in~\cite{van2022newinla}. The concepts discussed in this work are equally applicable to the two formulations. In both cases we assume that $\vec{x}$ forms a Gaussian Markov random field (GMRF) with zero mean and sparse precision matrix $\Q(\th)$, i.e. $\x \sim \mathcal{N}(0, \Q(\th))$. One of the key components to INLA's computational efficiency is this sparse GMRF structure. 
It allows for a very high-dimensional parameter space without performance
degradation due to the employed sparse linear algebra
operations~\cite{rue2005gaussian}. 
In addition one adopts a prior $\pi(\th)$ for the hyperparameters $\th$. 
Thus, forming a three-stage model consisting of the hyperparamter distribution, the latent field
and the likelihood. 
It is also possible account for constraints imposed on the latent parameter space of the model~\cite{rue2005gaussian}. For simplicity, we will not discuss this case in further detail, as it is similar to the unconstrained framework when it comes to parallelization strategies. 
For a further discussion of the model specifications or Bayesian hierarchical models
see~\cite{rue2017bayesian, congdon2014applied}.

% TODO: find better subheading
\subsection{Approximating $\pi(\theta_j \vert
    y)$} % or alternatively \pi(\theta \vert y)
\label{subsecApproxPostTheta}
\begin{comment}
    ONE POSSIBILITY : say we limit ourselves to Gaussian case as
    parallelisation scheme is basically the same in the two cases -
    important for me generally keep in mind: for non-Gaussian case,
    there is an inner iteration in \pi_G(x | \th, y) (details later),
    usually 2-3 iterations necessary, additionally optimisation
    procedure (also keep in mind that this is NOT necessary for the
    prior \pi(\x | \th), hence the parallelisation that i put there in
    my code will not help so much ... )
\end{comment}

% introduce posterior hyperparamters
INLA does not provide an estimate to the full joint posterior
$\pi(\x, \th \vert \y)$ but instead computes the posterior marginal
distributions $\pi(\theta_j \vert \y)$ and $\pi(x_i \vert \y)$ as well
as other relevant statistics~\cite{rue2009approximate}. This is done
through a nested approximation scheme that makes use of the following
estimate. The posterior of the hyperparamters $\th$, given the
observations, is approximated using Bayes' rule where each term in the
quotient is considered separately as follows
% write out posterior of hyperparamters
\begin{equation}
    \begin{aligned}
        \pi(\th\vert\vec{y}) &= \frac{\pi(\vec{x},\th\vert\vec{y})}{\pi(\vec{x}\vert\th, \vec{y})} \propto \frac{\pi(\th)\pi(\vec{x}\vert\th)\pi(\vec{y}\vert\vec{x},\th)}{\pi(\vec{x}\vert\th, \vec{y})}  \\
        &\approx
        \frac{\pi(\th)\pi(\vec{x}\vert\th)\pi(\vec{y}\vert\vec{x},\th)}{\pi_G(\vec{x}\vert\th,
            \vec{y})} \bigg\vert_{\vec{x} = \vec{x}^*(\th)} :=
        \tilde{\pi}(\th \vert \y).
        \label{eq:post_theta_approx}
    \end{aligned}
\end{equation}
The term $\pi_G(\x \vert \th, \y)$ describes a Gaussian approximation
of $\pi(\x \vert \th, \y)$ evaluated at the mode of $\x$ for a given
$\th$.
% use posterior hyperparamters around its mode for marginals
Through an iterative optimization algorithm the hyperparameter
configuration $\th^*$ for which
$\th^* = \operatorname*{arg\,max} \ \tilde{\pi}(\th \vert \y)$ is
found. Once the maximum is determined, the area around the mode is
explored and a set of selected evaluation points $\{\th^k\}_{k=1}^K$
is chosen, at which $\tilde{\pi}(\th^k \vert \y)$
% = \tilde{\pi}(\th^k \vert \y)$
is evaluated to approximate the posterior marginals for each
hyperparameter $\theta_j$ using numerical integration.
% For clarity of notation we want to point out here, that
% $\th^k$ refers to a vector of the same dimension as
% $\th$, denoting a particular hyperparameter configuration. On the
% hand $\theta_j$ refers to the
% $j$-th component, and thus $\tilde{\pi}(\theta_j \vert
% \y)$ describes the approximation to the posterior marginal
% distribution of the $j$-th parameter.
There is no global information available on $\tilde{\pi}(\th \vert
\y)$, instead it can only be evaluated for fixed values of
$\th$. Each function evaluation is, however, a computationally
expensive operation, mainly for two reasons. Let us for simplicity
first consider the case where the likelihood is Gaussian. To evaluate
the different terms of Eq.~(\ref{eq:post_theta_approx}) for a given
$\th$, we require the normalizing constant of the prior of the latent
effects $\pi(\x \vert \th)$. As mentioned previously,
$\x$ forms a GMRF, and hence the normalizing constant is computed from
the determinant of its precision matrix. The computational complexity
of computing the determinant of a dense matrix is in general
$O(n^3)$, where
$n$ denotes the size of the matrix~\cite{solomon2015numerical}. For
sparse matrices this cost can be lowered through employing specialized
solution methods but it nevertheless continues to pose an expensive
computational task for high-dimensional latent parameter
spaces~\cite{davis2006direct}.
% $\vec{Q}$
% This is done through a Cholesky factorization $\vec{L} \vec{L}^T =
% \vec{Q}$. In the case that the precision matrix
% $\vec{Q}$ is high-dimensional, it is crucial to have a performant
% library to carry out the required sparse computations, which we will
% discuss in detail in Sec.~\ref{subsecSparseLinAlg}.
To evaluate $\pi_G(\x \vert \th, \y)$ for a given
$\th$, one requires its normalizing constant as well as its mode which
is dependent on
$\th$. This requires the computation of the determinant and solving a
sparse linear system of size $n \times
n$, respectively, which are again computationally expensive tasks.
Fortunately, the sparsity pattern of the precision matrix of
$\pi_G$ is maintained when conditioning on the data
$\y$ which is an essential ingredient to the scalability of the INLA
methodology~\cite{rue2009approximate}. Determining the remaining terms
of Eq.~(\ref{eq:post_theta_approx}) is comparatively cheap from a
computational perspective. When considering the case, where the
likelihood is non-Gaussian, we have to perform an inner or nested
optimization, to obtain a good approximation $\pi_G(\x \vert \th,
\y)$ \cite{rue2009approximate}. This requires the repeated evaluation
of its normalizing constant and mode during each iteration, thus
further increasing the computational cost.
% As operations also vary depending on the model type, more details
% will be discussed in Sec.~\ref{sec:applications} using the concrete
% models at hand. potentially name it something like Q_x or whatever
% only cover simplified Laplace approximation

% TODO: better subsection header
\subsection{Approximating $\pi(x_i \vert y)$}
\label{subsecApproxMargLatP}
The posterior marginals $\pi(x_i \vert \y)$ can be written as
\begin{align}
  \pi(x_i \vert \y) = \int \pi(\th \vert \y) \pi(x_i \vert \th, \y) d\th,
  \label{eq:LatParam}
\end{align}
for which an approximation to the first factor of the right-hand side
and numerical integration points
$\{\th^k\}_{k=1}^K$ are already known from the previous step. It
remains to estimate $\pi(x_i \vert \th,
\y)$. If the likelihood is normally distributed, the Gaussian
approximation $\pi_G(\x \vert \th,
\y)$ from Eq.~(\ref{eq:post_theta_approx}) is exact. The posterior
marginals of $\pi(x_i \vert \th,
\y)$ can then be approximated by extracting the latent parameter
configuration $\x$ at the mode of $\pi(\th \vert
\y)$ from which one can directly infer the marginal means $x_i \vert
\x_{-i}, \th, \y$. Here, $\x_{-i}$ denotes the vector
$\x$ without the
$i$-th entry. The marginal variances are computed from
$\pi_G$, again, evaluated at the modal configuration. This, however,
requires at least a partial inversion of the precision matrix of
$\pi_G$, a computationally intensive operation especially for
high-dimensional latent parameters spaces.
% , which we will discuss in more detail in Sec.~\ref{subSecPartInv}.
In the non-Gaussian case, the approximation procedure is more
involved. An update for the correction of the conditional mean of $x_i
\vert \x_{-i}, \th, \y$ for each
$i$ needs to be performed. Additionally the marginal variances are
computed at all integration points
$\{\th^k\}_{k=1}^K$ which entails
$K$ partial matrix inversions. This allows INLA to more accurately
account for the overall shape and potential skewness in the
distributions. Using this information an approximation to the marginal
distributions $\pi(x_i \vert \th,
\y)$ are constructed, which are then used in Eq.~(\ref{eq:LatParam})
to compute the posterior marginals using numerical integration. For a
detailed overview we refer to~\cite{rue2009approximate}.
% write out posterior marginal of latent parameters what exactly is
% the parallelization scheme there?
Generally all computations are performed in log-scale as this is
favorable for dependence structures and in terms of numerical
robustness.

% talk about function evaluation, talk about sparse linear algebra

\begin{comment}
    - selected inversion is computed in every previously chosen
    integration point -> we can compute as many in parallel as we have
    threads (usually tied to the dimension of the the gradient, but
    clearly this is also linked to number integration points) - then
    weighted average of this is used for the marginal distributions -
    fitting spline, update mean, etc. not so important here
\end{comment}

% TODO: better subsection title
\subsection{Sparse Linear Algebra}\label{subsecSparseLinAlg}
% TODO: I have really no clue how much i can assume to be "known" and
% what is non-trivial

In this section we want to provide a quick overview of the underlying
sparse linear algebra techniques leveraged by the R-INLA package. A
sparse matrix is simply defined as a matrix where the majority of its
elements are equal to zero. For sparse matrix algebra it is customary
to only save the non-zero entries of the matrix in so-called
compressed matrix formats, that store the non-zero values and their
corresponding position instead of large quantities of zero entries.
This allows to massively reduce the required memory and computations
but hence, also demands for solvers that are especially tailored to
sparse problems. These arise, however, very commonly and are therefore
extensively researched, for introductory purposes
see~\cite{davis2006direct,saad2003iterative}.

During the first stage of the INLA algorithm, see
Sec.~\ref{subsecApproxPostTheta}, the log determinant of different
precision matrices $\Q$ (to obtain the corresponding normalizing
constant) and the solution to linear systems of equations of the form
$\Q \x=\vec{b}$ (to obtain the mode of the associated normal
distribution) are required. Since the arising precision matrices are
symmetric positive-(semi)definite, it is most efficient to perform a
Cholesky decomposition, where the original matrix $\Q$ factors into
$\Q = \vec{L} \vec{L}^T$, with $\vec{L}$ being a lower-triangular
matrix~\cite{ascher2011first}. The log determinant of $\Q$ can easily
be computed from the diagonal entries of $\vec{L}$ as
$\sum_{i=1}^{n} 2 \ \text{log} (L_{ii})$ and the system
$\Q \x = \vec{b}$ can quickly be solved for $\x$ using
forward-backward substitution. That means first solving
$\vec{L} \vec{y} = \vec{b}$ for $\vec{y}$ and then
$ \vec{L}^T \x = \vec{y}$ for $\x$. These operations can be executed
efficiently due to the lower-triangular structure of $\vec{L}$.

% \textcolor{red}{Say what is in old INLA version.}

\subsubsection{Matrix Reordering}
\label{subsecSymFact}
% TODO: emphasize parallelisation aspect: its not only important to
% find the permutation with the lowest fill-in but keep in mind the
% load balancing so that we can recursively break down the problem in
% smaller and smaller subproblems which we then solve independently &
% in parallel -> then we add back in the first sets of cut edges,
% potentially permute there, update, go to next level of cut edges

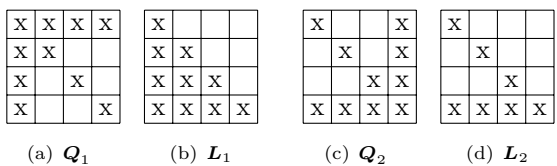
\begin{figure}[h!]
    \centering \subfigure[$\Q_1$]{
        \begin{tikzpicture}[scale=0.75, every node/.style={minimum
                size=.5cm-\pgflinewidth, outer sep=0pt}]
            \draw[step=0.5cm,color=black] (-1,-1) grid (1,1);
            % diagonal
            \node at (-0.75,+0.75) {x}; \node at (-0.25,+0.25) {x};
            \node at (+0.25,-0.25) {x}; \node at (+0.75,-0.75) {x};
            % arrow
            \node at (-0.25,+0.75) {x}; \node at (+0.25,+0.75) {x};
            \node at (+0.75,+0.75) {x}; \node at (-0.75,+0.25) {x};
            \node at (-0.75,-0.25) {x}; \node at (-0.75,-0.75) {x};
        \end{tikzpicture}} \subfigure[$\L_1$]{
        \begin{tikzpicture}[scale=0.75, every node/.style={minimum
                size=.5cm-\pgflinewidth, outer sep=0pt}]
            \draw[step=0.5cm,color=black] (-1,-1) grid (1,1);
            % diagonal
            \node at (-0.75,+0.75) {x}; \node at (-0.25,+0.25) {x};
            \node at (+0.25,-0.25) {x}; \node at (+0.75,-0.75) {x};
            % arrow
            \node at (-0.25,-0.75) {x}; \node at (-0.25,-0.25) {x};
            \node at (+0.25,-0.75) {x}; \node at (-0.75,+0.25) {x};
            \node at (-0.75,-0.25) {x}; \node at (-0.75,-0.75) {x};
        \end{tikzpicture}
    } \hspace{2pt} \subfigure[$\Q_2$]{
        \begin{tikzpicture}[scale=0.75, every node/.style={minimum
                size=.5cm-\pgflinewidth, outer sep=0pt}]
            \draw[step=0.5cm,color=black] (-1,-1) grid (1,1);
            % diagonal
            \node at (-0.75,+0.75) {x}; \node at (-0.25,+0.25) {x};
            \node at (+0.25,-0.25) {x}; \node at (+0.75,-0.75) {x};
            % arrow
            \node at (-0.25,-0.75) {x}; \node at (+0.75,+0.75) {x};
            \node at (+0.25,-0.75) {x}; \node at (+0.75,+0.25) {x};
            \node at (+0.75,-0.25) {x}; \node at (-0.75,-0.75) {x};
        \end{tikzpicture}} \subfigure[$\L_2$]{
        \begin{tikzpicture}[scale=0.75, every node/.style={minimum
                size=.5cm-\pgflinewidth, outer sep=0pt}]
            \draw[step=0.5cm,color=black] (-1,-1) grid (1,1);
            % diagonal
            \node at (-0.75,+0.75) {x}; \node at (-0.25,+0.25) {x};
            \node at (+0.25,-0.25) {x}; \node at (+0.75,-0.75) {x};
            % arrow
            \node at (-0.25,-0.75) {x};
            % \node at (-0.25,-0.25) {x};
            \node at (+0.25,-0.75) {x};
            % \node at (-0.75,+0.25) {x};
            % \node at (-0.75,-0.25) {x};
            \node at (-0.75,-0.75) {x};
        \end{tikzpicture}
    }
    \caption{Sparsity patterns of the four-by-four symmetric
        positive-definite arrowhead matrices $\Q_1$ and $\Q_2$, where
        each x stands for a non-zero entry and $\Q_2$ is a symmetric
        permutation of $\Q_1$. The sparsity patterns of their Cholesky
        factors are represented by $\L_1$ and $\L_2$, respectively.
        The matrix $\L_1$ is dense i.e. exhibits a large fill-in while
        $\L_2$ preserves the sparsity pattern of $\Q_2$.}
    \label{fig:arrowhead_matrices}
\end{figure}

Additionally it is in most cases advantageous to apply a symmetric
permutation to the matrix $\Q$ (and hence $\vec{b}$, respectively)
before computing the Cholesky decomposition. Even if the matrix $\Q$
is very sparse, $\L$ can have a large fill-in, meaning that there are
many entries which are non-zero in $\L$ but equal to zero in $\Q$. In
the elimination process that is used to compute $\L$, similar to
classical Gaussian elimination, one can see how these non-zeros arise
\cite{davis2006direct}. As an illustrative example, we consider a
symmetric positive-definite arrowhead matrix $\Q_1$ whose sparsity
pattern can be seen in Fig.~\ref{fig:arrowhead_matrices}. Its
corresponding Cholesky factor $\L_1$ is computed recursively, starting
from the first diagonal entry. % $L_{11}$ = \sqrt{Q_{11}}
\begin{comment}
    , and more generally we have for diagonal and off-diagonal entries
    of $\L$ that
    \begin{align}
      L_{jj} &= \sqrt{Q_{jj} - \sum_{k=1}^{j-1} L^2_{jk}}, \\
      L_{ij} &= \frac{1}{L_{jj}} \left( Q_{ij} - \sum_{k=1}^{j-1} L_{ik} L_{jk} \right) \ \text{for } i > j.
    \end{align}
\end{comment}
To compute the off-diagonal entry $(\L_1)_{32}$, we use that
$(\L_1)_{32}=((\Q_1)_{32}-(\L_1)_{31}(\L_1)_{21})/(\L_1)_{22}$. Hence,
even if $(\Q_1)_{32} = 0$, $(\L_1)_{32}$ is not equal to zero unless
$(\L_1)_{31}$ or $(\L_1)_{21}$ are. More generally one can observe a
dependency structure on previous columns of the $i$-th and $j$-th row.
By applying a symmetric permutation to $\Q_1$, we can obtain $\Q_2$.
If we now compute $\L_2$ we can see that $(\L_2)_{32} = 0$, since
$(\Q_2)_{32}, (\L_2)_{21}$ and $(\L_2)_{32}$ are all equal to zero.

In general, fill-in can often be drastically reduced by finding a
suitable matrix reordering of $\Q$ which then lowers the overall
memory and computation requirements of $\L$.
% Hence, an important objective of the matrix reordering is to reduce
% the expected fill-in of the factor $\vec{L}$ to lower the overall
% memory and computation requirements. This reduces the and is
% therefore beneficial.

For each of the different arising precision matrices in INLA the
sparsity pattern never changes throughout the optimization phase but
only the numerical values of the non-zero entries. Hence, it is
sufficient to only compute suitable reorderings once throughout the
entire algorithm. Finding a favorable permutation is, however, a
challenging task and has hence been, an active area of
research~\cite{heath1991parallel, bichot2013graph}.

\begin{figure*}[ht]%
    \centering \subfigure[]{ \fbox{\begin{tikzpicture}[scale=3]
                \foreach \i in {0,...,\nrows}{ \foreach \j in
                    {0,...,\ncols}{ \pgfplotstablegetelem{\i}{\j}\of\A
                        \ifnum\pgfplotsretval=0\relax\else
                        \node[rectangle, draw=black, minimum size=3pt,
                        inner sep=0pt,
                        fill=mylightblue!\pgfplotsretval!mylightblue]
                        at (\j pt,-\i pt) {}; \fi }; };
            \end{tikzpicture}} % end fbox
    } % end subfigure
    % chol(A)
    \subfigure[]{ \fbox{\begin{tikzpicture}[ scale=3] \foreach \i in
                {0,...,\nrows}{ \foreach \j in {0,...,\ncols}{
                        \pgfplotstablegetelem{\i}{\j}\of\cholA
                        \ifnum\pgfplotsretval=0\relax\else
                        \node[rectangle, draw=black, minimum size=3pt,
                        inner sep=0pt,
                        fill=mylightblue!\pgfplotsretval!mylightblue]
                        at (\j pt,-\i pt) {}; \fi }; };
            \end{tikzpicture}} % end fbox
    } % end subfigure
    \hspace{10pt}
    % A permuted
    \subfigure[]{ \fbox{\begin{tikzpicture}[scale=3] \foreach \i in
                {0,...,\nrows}{ \foreach \j in {0,...,\ncols}{
                        \pgfplotstablegetelem{\i}{\j}\of\permA
                        \ifnum\pgfplotsretval=0\relax\else
                        \node[rectangle, draw=black, minimum size=3pt,
                        inner sep=0pt,
                        fill=mylightblue!\pgfplotsretval!mylightblue]
                        at (\j pt,-\i pt) {}; \fi }; };
                                
                % add indicators for submatrices
                \node[rectangle, draw=black, minimum size=42,
                label={[xshift=0.4cm, yshift=-0.6cm]:$Q_A$}] at (6.5
                pt, -6.5 pt){}; \node[rectangle, draw=black, minimum
                size=42, label={[xshift=0.4cm, yshift=-0.6cm]:$Q_B$}]
                at (20.5 pt, -20.5 pt){}; \node[rectangle, draw=black,
                minimum width = 90, minimum height = 4,
                label={[xshift=-1cm, yshift=0cm]:$Q_S$}] at (14.5 pt,
                -28.5 pt){};
                                
            \end{tikzpicture}} } % end subfigure
    % chol(A_permuted)
    \subfigure[]{ \fbox{\begin{tikzpicture}[scale=3] \foreach \i in
                {0,...,\nrows}{ \foreach \j in {0,...,\ncols}{
                        \pgfplotstablegetelem{\i}{\j}\of\permCholA
                        \ifnum\pgfplotsretval=0\relax\else
                        \node[rectangle, draw=black, minimum size=3pt,
                        inner sep=0pt,
                        fill=mylightblue!\pgfplotsretval!mylightblue]
                        at (\j pt,-\i pt) {}; \fi }; };
                                
                % add indicators for submatrices
                \node[rectangle, draw=black, minimum size=42,
                label={[xshift=0.4cm, yshift=-0.6cm]:$L_A$}] at (6.5
                pt, -6.5 pt){}; \node[rectangle, draw=black, minimum
                size=42, label={[xshift=0.4cm, yshift=-0.6cm]:$L_B$}]
                at (20.5 pt, -20.5 pt){}; \node[rectangle, draw=black,
                minimum width = 90, minimum height = 4,
                label={[xshift=-1cm, yshift=0cm]:$L_S$}] at (14.5 pt,
                -28.5 pt){};
                                
            \end{tikzpicture}} } % end subfigure
        \caption{From left to right: (a) Lower triangular part of the sparsity pattern of a random symmetric positive definite matrix $\vec{Q}$, with 70 non-zero entries. (b) Cholesky decomposition $\vec{L}$ of $\vec{Q}$, where $L$ has 180 non-zero entries. (c) Permuted matrix $\vec{Q}_{\text{perm}}$ using (nested) dissection. (d) Cholesky decomposition $\vec{L}_{\text{perm}}$ of $\vec{Q}_{\text{perm}}$, where $\vec{L}_{\text{perm}}$ has only 88 non-zero entries. We can observe that $\vec{Q}$ was permuted such that it is separated in two independent submatrices $\Q_A$ and $\Q_B$, which are only connected through entries that are now placed in the two rows, which we refer to as $\Q_S$. This way there is no fill-in between $\Q_A$ and $\Q_B$ but only within the before-mentioned submatrices. Thus, reducing the number of non-zeros in the Cholesky factorization of $\Q_{\text{perm}}$.
                %This way the two submatrices can be factorized parallel, i.e. we can compute $\L_A$ and $\L_B$ in parallel, before computing the entries of $\L_S$. Hence, there are not only less entries to compute in $\vec{L}_{\text{perm}}$ compared to $\L$, but a large part of the computation can also be done in parallel. 
            The strategy can be applied recursively to each of the
            components $\Q_A$ and $\Q_B$. }
        \label{fig:matrix_perm}
    \end{figure*}
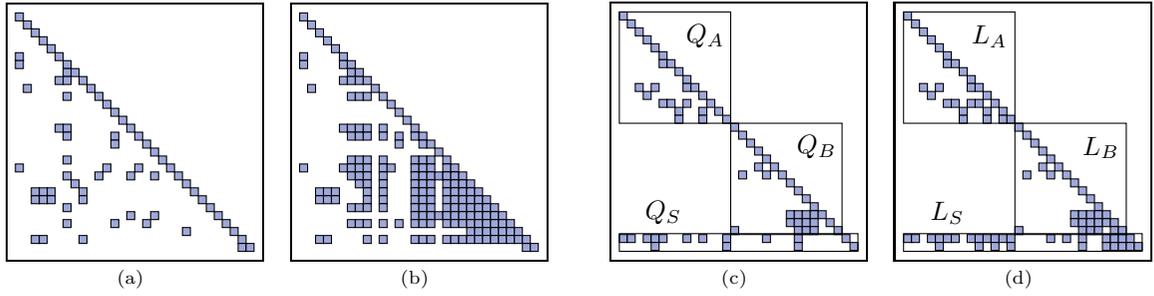

A simple greedy strategy is the minimum degree ordering, where columns
are successively permuted to minimize the number of non-zero
off-diagonal entries of the current pivot element
\cite{george1989evolution}. Another popular class of reordering
strategies called nested dissection employs graph partitioning
techniques \cite{george1973nested, bichot2013graph}. A symmetric
matrix can be associated with an undirected graph, where each diagonal
matrix entry represents a node and each nonzero off-diagonal entry an
edge between the corresponding nodes. The graph is partitioned into
two roughly equal-sized independent subgraphs
$\mathcal{G}_A, \mathcal{G}_B$ which are only connected through a
separating set $\mathcal{G}_S$. The partition is chosen such that the
size of the separating set is minimized. For the corresponding matrix
this means that the associated submatrices $\Q_A$ and $\Q_B$ can be
written as block matrices only connected through $\Q_S$. Hence, all
fill-in that can occur is within $\Q_A, \Q_B$ and $\Q_S$, see
Fig.~\ref{fig:matrix_perm}.
    For large graphs the computational cost is often too high to
    compute an optimal separating set. Therefore, multilevel
    strategies are used to recursively coarsen a large graph by
    merging connected vertices until it is reduced to a tractable
    size. For the coarse graph an edge cut can be defined which is
    then recursively refined while being propagated back to the
    original large graph. From the final edge cut we can deduce a
    separating set $\mathcal{G}_S$ \cite{karypis1998fast}. The same
    strategy of finding roughly balanced minimal separating sets can
    be independently reapplied to $\Q_A$ and $\Q_B$, leading to the
    nested dissection approach.
    % Depending on the particular structure of $\Q$, no guarantee for
    % improvements, etc. TODO: @olaf : additional ideas what to cite?
    % For a comprehensive overview we refer to \cite{bichot2013graph}.
    A popular state-of-the-art library that computes such heuristic
    matrix reorderings is called METIS \cite{karypis1998fast} and is
    used by both, TAUCS as well as PARDISO. While the previously
    employed TAUCS library is limited to using fill-in reducing
    permutations, PARDISO can additionally perform much of the
    Cholesky factorization in parallel, making use of the independent
    submatrices arising from the nested dissection reordering. This
    will be explained in detail in Sec.~\ref{subSecParPARDISO}.

\subsubsection{Partial Inversion}
\label{subSecPartInv}

\begin{comment}
    Two ways to explain this: using graphs and matrix identities,
    takahashi or using conditional indepence properties (which boils
    down to graphs as well but ok) we can derive entries \Sigma_ij
    also completely from that -> probably the way to go! check Rue &
    Martino 2007 or handbook of spatial statistics chapter 12
\end{comment}

\begin{comment}
    for my own understanding: $\pi(x_i \vert \y)$ is a non-necessarily
    Gaussian distribution. We assume it to be unimodal and estimate
    the mean somehow. To estimate the variance we use the evaluation
    points in theta space and estimate for these different points the
    partial inverses. Then we somehow assemble these partial inverses
    to an overall distribution for each point. I guess this is where
    the story of the splines comes in? Because so far we only have
    individual approximations to the variance in theta space.
\end{comment}

In the second stage of the INLA algorithm, see
Sec.~\ref{subsecApproxMargLatP}, an estimate for the variances of the
posterior marginals of the latent parameters, i.e. $\pi(x_i \vert \y)$
for all $i$, is obtained. To do so INLA uses information from the
Gaussian approximations $\pi_G(\x \vert \th^k, \y)$ at the various
integration points $\{\th^k\}_{k=1}^K$. All information about the
marginal variances only exists, however, in form of the precision
matrices.
% TODO: Introduce conditional independence properties etc. as way to
% extract variance
As the variances $\Sigma_{ii}$ correspond to $(\Q^{-1})_{ii}$, they
can only be obtained through matrix inversions. If the full precision
matrices had to be inverted this would become very time and memory
consuming if not infeasible for high-dimensional latent parameter
spaces as matrix inversions have a complexity of $O(n^3)$ for a matrix
of size $n \times n$.
\begin{comment}
    For the Gaussian case: We compute
    $\Q_{\x \vert \th, y} at the mode of $\th$ for $\pi(\th \vert
    \y)$. Then we invert. This gives us the variances of the latent
    parameters. For the non-Gaussian case: we compute the inverses at
    all integration points and puzzle them together.
\end{comment}
% The inverse of a sparse matrix is in general dense.
%%%% write statistics approach to this.
Fortunately there is an alternative method for computing the marginal
variances that is much more efficient. There are two different
approaches to derive this recursive strategy. One is using the
conditional independence properties of GMRFs and their associated
graphical structure, and was first described by Rue and Martino in
\cite{rue2007approximate}. The other one is based on a particular way
of writing matrix identities without a further interpretation but
describing the same recursion and was developed by
Takahashi~\cite{takahashi1973formation} almost 50 years ago. We will
begin by considering the former. The solution
$\x$ to the problem $\vec{L}^T \x = \vec{z}$ where $\vec{z} \sim
\mathcal{N}(\vec{0},
\vec{I})$ is a sample from a GMRF with zero mean and precision matrix
$\Q = \L \L^T$ \cite{rue2005gaussian}.
% conditional distribution
Since
$\vec{L}^T$ is an upper triangular matrix, we can use a backward solve
and recursively compute
\begin{equation}
    \begin{aligned}
        x_n &= \frac{z_n}{L_{nn} } \\
        x_i &= \frac{z_i}{L_{ii}} - \frac{1}{L_{ii}}
        \sum_{\substack{k>i \\ L_{ki} \neq 0}}^n L_{ki} x_k
    \end{aligned}
    \label{eq:backward_solve}
\end{equation}
for $i=n-1, ..., 1$. We exclude all terms
$L_{ki}$ from the summation directly that equate to zero. This way it
becomes clear that the more zeros we have in
$\L$, the less computations are necessary. If we use the known
expected value and variance of
$\vec{z}$ as well as Eq.~(\ref{eq:backward_solve}), we can deduce the
covariance matrix $\vec{\Sigma}$ of
$\x$ from first principles, obtaining
\begin{equation}
    \Sigma_{ij} = \frac{\delta_{ij}}{L_{ii}^2} - \frac{1}{L_{ii}} \sum_{\substack{k>i \\ L_{ki} \neq 0}}^n L_{ki} \Sigma_{kj},
    \label{eq:sel_inv_cov}
\end{equation}
where $\delta_{ij}$ is one if $i =
j$ and zero otherwise. The entries can only be computed recursively
starting at
$i=n$, traversing the matrix from the bottom right to the top left.
% For a sparse factor
% $\L$, this can dramatically reduce the required computations

% TODO: Is this section too detailed?
Equivalently we can derive Eq.~(\ref{eq:sel_inv_cov}) without using
statistical properties. Instead we consider the slightly altered
decomposition $\Q = \L \L^T = \vec{V} \vec{D} \vec{V}^T$, i.e. $\L =
\vec{V} \vec{D}^{1/2}$, where
$\vec{D}$ is a diagonal matrix and
$\vec{V}$ a lower triangular matrix with ones on the diagonal. The
following matrix identity was proposed by Takahashi in
\cite{takahashi1973formation},
\begin{equation}
    \vec{\Sigma} = \vec{D}^{-1}\vec{V}^{-1} + (\vec{I} - \vec{V}^T)\vec{\Sigma}.
    \label{eq:takahashi_identity}
\end{equation}
As
$\vec{\Sigma}$ is a symmetric matrix, it is enough to compute its
upper triangular part. The term
$\vec{D}^{-1}\vec{V}^{-1}$ is lower triangular with
$(\vec{D}^{-1}\vec{V}^{-1})_{ii} =
(\vec{D}^{-1})_{ii}$, since
$\vec{V}$ has a unit diagonal. Writing out
Eq.~(\ref{eq:takahashi_identity}) as sums we obtain
\begin{equation}
    \begin{aligned}
        \Sigma_{ij} = \frac{\delta_{ij}}{D_{ii}} - \sum_{\substack{k>i
                \\ V_{ki} \neq 0}} V_{ki} \Sigma_{kj}, \quad
        \text{for} \ i \leq j
    \end{aligned}
    \label{eq:takahashi_recursion}
\end{equation}
which we can compute recursively starting from
$\Sigma_{nn}$ and where again, $\delta_{ij}$ is equal to one for
$i=j$ and zero otherwise. Eq.~(\ref{eq:sel_inv_cov}) and
(\ref{eq:takahashi_recursion}) are equal since $\L = \vec{V}
\vec{D}^{1/2}$. It is possible to use these recursions to compute all
entries of
$\vec{\Sigma}$, however, in this case they do not give us a
computational advantage over traditional inversion algorithms. In both
cases we have a complexity of
$O(n^3)$. If we are instead only interested in particular entries of
$\vec{\Sigma}$, e.g. only the diagonal, and if additionally
$\L$ (and then likewise
$\vec{V}$) are sparse, a tremendous amount of computational cost can
be saved, as only the entries $\Sigma_{ij}$ for which
$L_{ij}$ is non-zero need to be computed.
% We can see that in order to compute the marginal variances, we do
% not have to compute all entries of
% $\vec{\Sigma}$ but only those $\Sigma_{ij}$ for which
% $L_{ij}$ is non-zero. For a more detailed analysis we refer to
% \cite{rue2007approximate}. TODO: 1 longer sentence why?
% \textcolor{red}{Considering the permuted Cholesky decomposition of
% the example matrix from Fig.~\ref{fig:matrix_perm}(d) we have ...
% non-zeros which will lead to ... a ... } Fortunately there are
% routines that allow us to only compute specific entries of the
% inverse. Namely those who have non-zero values in their Cholesky
% factor
% $\vec{L}$. takahashi One of those algorithms was originally
% developed by Takahashi~\cite{takahashi1973formation} almost 50 years
% ago. Cleverly rewriting $\Q \vec{\vec{\Sigma}} =
% \vec{I}$, using different matrix identities, we can recursively
% solve for
% $\vec{\Sigma}$. While it is possible to compute all entries of
% $\vec{\Sigma}$ which is by definition
% $\Q^{-1}$, we can limit ourselves to computing only the entries
% $(\vec{Q}^{-1})_{ij}$ for all $i,j$ where $\vec{L}_{ij} \neq
% 0$. It is worth noting that since
% $\Q$ is symmetric, so is its inverse and hence it is enough to only
% consider $\vec{L}$ and not also its transpose.
Hence, the less non-zeros we have in the factor
$\vec{L}$, the smaller is the computational need. This highlights the
importance of finding suitable permutations as described in the
previous section.

PARDISO and INLA's previous selected inversion routine both employ
such a partial inversion strategy. While the latter is sequential,
PARDISO is using parallelized computations for a shorter time to
solution, exploiting the particular matrix structure given through the
permutation, see Sec.~\ref{subSecParPARDISO} for details. The marginal
variances $\Sigma_{ii}$ that we obtain are then repermuted to
correspond to the original ordering of the parameters $\x$.
Afterwards, they can be used as described in
Sec.~\ref{subsecApproxMargLatP} to obtain estimates of the marginal
posterior distributions of the latent parameters.

\begin{comment}
    \begin{equation}
        x_i \vert x_{i+1}, ..., x_n \sim \mathcal{N} \left( - \frac{1}{L_{ii}} \sum_{j=i+1}^n L_{ji} x_j, \frac{1}{L_{ii}^2} \right)
    \end{equation}
\end{comment}

\section{New Parallelization Scheme}\label{sec:parallel_scheme}

%% motivation motivation
In the following we present the novel parallelization strategies that
have been added to the current R-INLA implementation. There are four
significant updates in the algorithm that we want to discuss in this
work. The first two comprise parallelized versions of the gradient
computations and the posterior marginal variance estimations.
Additionally, we introduced a parallel line search strategy within the
optimization routine of the algorithm. The parallelizations are
realized through the usage of OpenMP, an application programming
interface that supports multiprocessing for shared-memory
architectures and allows to efficiently execute multiple tasks at
once. The last update applies to all stages of the algorithm and
consists of the incorporation of the PARDISO library into R-INLA to
handle all required major sparse linear algebra operations.

\subsection{Parallelization of Function Evaluations}
\label{subSecParGradient}
% / Function evaluations TODO: distinguish between f(\theta) as the
% whole function and f(\theta) as in f evaluated at \theta?! TODO:
% computation of partial inverse actually also parallelized add this
% as well !! , i.e. let $f(\th) = \tilde{\pi}(\th \vert \y)$ and then
% determine
% \begin{align}
%       \th^* = \operatorname*{arg\,max} f(\th). 
%\end{align}
Among the computationally intensive and therefore time-consuming
operations in INLA are the function evaluations of the posterior
marginal distribution, i.e. computing $\tilde{\pi}(\th \vert \y)$ at
$\th$. Instead of evaluating the posterior directly, we consider
\begin{equation}
    f(\th) := - \text{log} \ \tilde{\pi}(\th \vert \y).
\end{equation}
The logarithmic scale is introduced to enhance numeric stability. The
sign switch turns $f$ into a minimization problem to find the mode
$\th^*$ of the posterior.
% , named after their creators C. Broyden, R. Fletcher, D. Goldfarb
% and D. Shanno
To solve the optimization problem INLA uses a
BFGS-algorithm~\cite{nocedal2006numerical} which resembles a Newton
method, without requiring knowledge about the second order derivative.
Information on the gradient is, however, needed. We estimate it
numerically using a finite difference scheme~\cite{leveque2007finite}.
Each directional derivative $\partial f / \partial \theta_i$ in the
gradient is approximated using either a first order forward or central
difference.
% \textcolor{red}{Need to make sure that we are actually using
% directional derivative along the axes or not. But need to be
% consistent.}
As an example we show the central difference approximation along the
coordinate axes
\begin{equation}
    \frac{\partial f}{\partial \theta_i}(\th) \approx \frac{f(\th + \vec{\epsilon}_i) - f(\th - \vec{\epsilon}_i)}{2 \ \vert \vert \vec{\epsilon}_i \vert \vert} \ \text{for all } i.
    \label{eq:central_diff}
\end{equation}
The vector $\vec{\epsilon}_i$ is of the same dimension as $\th$ and
contains only zeros except for the $i$-th component which contains a
small value $\epsilon > 0$. Often the directional derivatives are
computed according to this canonical basis, however, non-canonical
bases can just as well be used. INLA makes use of knowledge from
previous iterations to choose directional derivatives exhibiting more
robust numerical properties and hence faster overall convergence, see
\cite{fattah2022smart} for details.
% We would like to explicitly point out the difference between the
% vector $\th_l$ and the partial derivative with respect to
% $\partial \theta_i$ to avoid confusion.
Independently of the choice of basis, the directional derivatives are
computed for each component of $\th$ and each time entail one or two
function evaluations of $f$.
% These are, as previously mentioned, computationally intensive due to
% the involved Cholesky decompositions.
We can, however, see from Eq.~(\ref{eq:central_diff}) that all
function evaluations are independent from each other. In the new
parallelization scheme we are using this to our advantage by computing
the function values, i.e.
$f(\th + \vec{\epsilon}_1), f(\th - \vec{\epsilon}_1), f(\th +
\vec{\epsilon}_2), ... $ simultaneously in each iteration.
% The dimension of the gradient is equal to the number of
% hyperparameters in the model.
For a central difference scheme we need two times as many function
evaluations as there are hyperparameters, which can now all computed
in parallel instead of sequentially, while also computing $f(\th)$
itself. So, if e.g. dim$(\th) = 3$ we can have a theoretical speedup
of 7 during the gradient computation. This means the introduced
parallelism allows the gradient to be computed at almost no further
cost in addition to the already necessary iterative evaluations of
$f(\th)$ for fixed values of $\th$.
% We will show in Sec.~\ref{sec:results} that this significantly
% decreases the overall runtime. Automatic differentiation
There are other methods to obtain gradient information, in particular
automatic differentiation~\cite{baydin2018automatic}. In the ideal
case it offers a highly accurate solution at runtimes comparable to a
single function evaluation~\cite{baydin2018automatic}, similar to the
parallelized finite difference approximation. It requires, however, an
adaptation of the basic linear algebra operations and then allows for
parallelism within these operations. This is not trivially done for
non-standard cases \cite{van2018automatic} and would require major
changes in the PARDISO software library. Additionally, if the
likelihood is non-Gaussian we have the inner optimization routine
within each function evaluation, for which we are not aware of
existing automatic differentiation methods.
% As we cannot analytically compute the gradient it needs to be
% determined otherwise.

After the mode of $\tilde{\pi}(\th \vert \y)$ is found, we can also
perform the function evaluations of the integration points
$\{\th^k\}_{k=1}^K$ in parallel. Additionally, we have to estimate the
posterior marginal variances, computing the partial inverses of the
precision matrices at the integration points, see
Sec.~\ref{subsecApproxMargLatP}. We also implemented this to be
executed in parallel, as these computations are also independent from
each other.

\subsection{Parallel Line Search}
\label{subSecParLineSearch}
% TODO: fix upper & lower case notation ...

\begin{figure}[h!]
    % \centering
    \begin{tikzpicture}[domain=-1.5:10, scale=0.57]
                                
        % draw axis layout
        \draw[help lines, color=gray!30, dashed] (-2,-0.5) grid
        (9.9,4.9); \draw[->,ultra thick] (-2.5,0)--(10,0)
        node[right]{$\theta$}; \draw[->,ultra thick] (-2,-0.5)--(-2,5)
        node[above]{$f(\theta)$};
                
        % \addplot {x^2-7*x+10} ;
        % \foreach \x in {2,3,...,9}
        % compute best fit polynomial using all data points ()
        % weighting scheme :
        \draw[color=black, thick] plot
        (\x,0.042*\x*\x-0.475*\x+3.8182) node[below]{$q$};
        % \draw[color=black, thick] plot (\x,0.048*\x*\x-0.55*\x+3.97)
        % node[below]{$q$};
                                                
        % draw \theta labels
        \draw[thick] (1,-0.2) -- (1, 0.2) node[below, midway,
        color=red, yshift=-1.5]{$\theta_l$}; \draw[thick] (9,-0.2) --
        (9, 0.2) node[below, midway,
        yshift=-1.5]{$\theta_l - \gamma_l p_l$};
                        
        % draw f(\theta)
        \draw (1,3.5) node[cross3,rotate=0] {}; \draw (1, 2.8)
        node[color=red]{\small $f(\theta_{l})$}; \draw (9,2.8)
        node[cross1,rotate=0] {};

        % draw additonal \theta from line search
        \draw[thick, color=gray!140, dashed] (-1,-0.2) -- (-1, 0.2)
        node[below, midway, yshift=-1.5]{$\theta_{l_{-2}}$};
        \draw[thick, color=gray!140, dashed] (0,-0.2) -- (0, 0.2)
        node[below, midway, yshift=-1.5]{$\theta_{l_{-1}}$};
        \draw[thick, color=gray!140, dashed] (3,-0.2) -- (3, 0.2)
        node[below, midway, yshift=-1.5]{$\theta_{l_1}$}; \draw[thick,
        color=gray!140, dashed] (5,-0.2) -- (5, 0.2) node[below,
        midway, yshift=-1.5]{$\theta_{l_2}$}; \draw[thick,
        color=gray!140, dashed] (7,-0.2) -- (7, 0.2) node[below,
        midway, yshift=-1.5]{$\theta_{l_3}$};
                        
        % draw corresponding f(theta) values
        \draw (3,2.7) node[cross2,rotate=0] {}; \draw (5,2.3)
        node[cross2,rotate=0] {}; \draw (7,2.9) node[cross2,rotate=0]
        {};
        % draw additional f(theta) to the left
        \draw (0,3.7) node[cross2,rotate=0] {}; \draw (-1,4.4)
        node[cross2,rotate=0] {};
                        
        % minimum
        \draw[thick] (5.77,-0.2) -- (5.77, 0.2) node[color=red, below,
        xshift=3, midway, yshift=-1.5]{$\theta_{l+1}$}; \draw
        (5.65,2.47) node[cross3,rotate=0] {}; \draw (6.2, 1.8)
        node[color=red]{\small $f(\theta_{l+1})$};
                        
        \draw (-1.1, 3.7) node[color=gray!]{\small
            $f(\theta_{l_{-2}})$}; \draw (3, 3.2)
        node[color=gray!]{\small $f(\theta_{l_1})$}; \draw (7, 3.4)
        node[color=gray!]{\small $f(\theta_{l_3})$};

    \end{tikzpicture}
    \caption{Illustration of the parallel line search using robust
        regression. Let $\th_l$ be the current iterate with function
        value $f(\th_l)$, new search direction $\vec{p}_l$, with
        search interval $I = (\th_l, \th_l - \gamma_l \vec{p}_l)$ on
        which points $\th_{l_i}$ are defined, with
        $\th_{l_0}:= \th_{l}$, and all $f(\th_{l_i})$ are evaluated in
        parallel. The evaluation points $\th_{l_{-1}}, \th_{l_{-2}}$
        in positive $\vec{p}_l$ direction are added for numerical
        stabilization. The polynomial $q$ is fitted using robust
        regression and its minimum becomes the next iterate
        $\th_{l+1}$.}
    \label{fig:parallel_linsearch}
\end{figure}
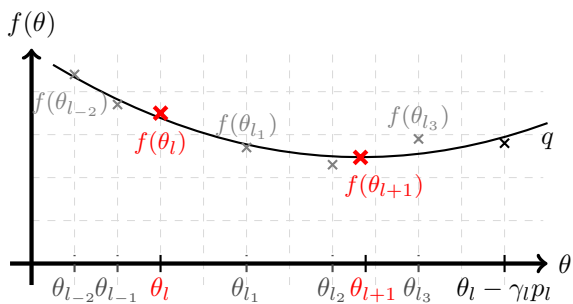

During the $l$-th iteration of the optimization routine
% which aims at locating the mode of
% $f(\th) = \tilde{\pi}(\th \vert \y)$,
we approximate the gradient $\nabla f_{\th_l}$ at $\th_l$.
% For a given $\th_l$, we determine the value of $f(\th_l)$ as well as
% an approximation to its gradient $\nabla f(\th_l)$, as described
% previously.
If a full Newton scheme was employed, one would additionally have to
compute the Hessian of $f$. When using a BFGS algorithm, an
approximation to the Hessian $\vec{B}_l$ is formed in each iteration,
utilizing information from the previous gradients. Each next iterate
is computed as
\begin{equation}
    \th_{l+1} = \th_l - \alpha_l [\vec{B}_l(f_{\th_l})]^{-1} \nabla f_{\th_l},
    \label{eq:bfgs_update}
\end{equation}
where $0 < \alpha_l < 1$ is the step size in iteration $l$.
Determining a suitable value for $\alpha_l$ is crucial to the
convergence of the overall algorithm and is done through a line search
procedure, see e.g. \cite{nocedal2006numerical} for an overview. The
most common approach is to use a mixture of artificially chosen upper
and lower bounds, as well as the value from the previous iteration, to
propose the next step size $\alpha_l$. Then, a check is performed to
see if a certain condition is met to either accept or reject the
suggested step size. If it is rejected, the step size is reduced, and
the check is performed again, if it is accepted, $\th_l$ is updated as
described in Eq.~(\ref{eq:bfgs_update}) and one proceeds to the next
iteration. For this check, the function $f$ is evaluated at the
potential new candidate which, as discussed previously, is an
expensive operation to perform, but necessary in order to determine
the validity of the new step. Hence, every time a step is rejected
another sequentially computed function evaluation is required. To be
more efficient we introduce a parallel line search strategy to make
better use of the available resources. Instead of computing different
values for $\alpha_l$ sequentially, we define a search interval
$I = (\th_l, \th_l - \gamma_l \vec{p}_l)$, where
$\vec{p}_l := [\vec{B}_l(f_{\th_l})]^{-1} \nabla f_{\th_l}$ and
$\gamma_l > 0$ is an upper bound to $\alpha_l$. Hence, $I$ contains
all possible solutions of Eq.~(\ref{eq:bfgs_update}) for
$0 < \alpha_l \leq \gamma_l$. Depending on the number of available
cores we define a number of points $\th_{l_i}$ on $I$ for which we
evaluate all $f(\th_{l_i})$ in parallel. The easiest option would be
to now choose the candidate $\th_{l_i}$ that minimizes $f$ as the next
iterate. However, it has shown to be advantageous to fit a second
order polynomial $q$ through the newly evaluated points using robust
regression \cite{rousseeuw2005robust, atkinson2000robust}. This way
slight inaccuracies in the function evaluations (that can e.g. arise
from more complicated choices of likelihood and require an inner
optimization loop) are counter balanced. We additionally add two more
evaluation points in positive $\vec{p}_l$ direction close to $\th_l$
to stabilize the polynomial fitting process close to the global
optimum. This does not increase the overall runtime as they can be
evaluated in parallel with the other $\th_{l_i}$. Robust regression
differs from regular regression in the sense that each pair
$(\th_{l_i}, f(\th_{l_i}))$ gets assigned a weight $w_{\th_{l_i}}$,
which will make it more or less influential on the overall fitting
process. There are a number of commonly used weighting functions $w$.
We have chosen to use the so-called bisquare weighting as this has
numerically shown to be most suitable.
% For the interested reader we refer e.g. to \cite{atkinson2000robust}
% for a further discussion on robust regression.
After finding the second order polynomial $q$ its minimum is
determined on $I$ and chosen as the next iterate $\th_{l+1}$. In
Fig.~\ref{fig:parallel_linsearch} we can see an illustration of the
new strategy. The advantages are the mitigation of inherently
sequential function evaluations in the reject-accept check, as well as
an improved step size choice. This can lower the number of required
iterations until convergence, and hence also the overall runtime, see
Sec.~\ref{sec:results} for numerical results.

\subsection{Parallelization within PARDISO}
\label{subSecParPARDISO}
% \textcolor{red}{use median not mean.}

% In this section we want to focus on the parallel strategies used
% within PARDISO.
The PARDISO library offers routines to efficiently provide solutions
to various problems commonly arising in the field of sparse linear
algebra~\cite{doi:10.1137/1.9781611971446}. In R-INLA the PARDISO
implementation for Cholesky decomposition, subsequent forward-backward
substitution and partial inversion are invoked.
% PARDISO is a state-of-the-art direct solver that has been in active
% development for many years and relies on various strategies for
% excellent performance. They range from the selection of algorithm to
% detailed hardware considerations.
PARDISO is a state-of-the-art direct solver that has been in active
development for many years. To achieve excellent performance it relies
on numerous strategies ranging from thorough algorithm selection to
detailed hardware considerations.

In this work we want to solely focus on the employed parallelism and
refer the interested reader to~\cite{bollhofer2020state} for a
comprehensive overview.

As described in Sec.~\ref{subsecSymFact}, a symmetric permutation
found through nested dissection is applied to $\Q$ before computing
its Cholesky decomposition $\L$. This reordering technique does not
only reduce the arising fill-in but also creates independent block
matrices $\Q_A, \Q_B$, see Fig.~\ref{fig:matrix_perm}, that are only
connected through a small submatrix $\Q_S$. Hence, it is possible to
factorize $\Q_A$ and $\Q_B$ in parallel, before factorizing $\Q_S$.
The reordering generated through nested dissection has a recursive
pattern, and thus $\Q_A$ and $\Q_B$ themselves exhibit the same
structure. Both of them, again, contain two independent blocks and a
connecting submatrix which corresponds to the separating set of the
associated graph. The reordered matrix can therefore be factorized in
parallel, starting from the set of smallest block matrices. If
sufficient compute resources are a available this introduction of
parallelism drastically reduces the computational time and especially
exhibits much better scaling. While it is not possible to give precise
bounds on timings without specifying more graph theoretical concepts
and notation, one can see that the recursive subdivisions induce a
logarithmic growth. For details see~\cite{pan1985efficient}.

A similar principle can be applied to determine the partial inverse of
$\Q$, for which we have seen in Sec.~\ref{subSecPartInv} that only the
non-zero entries of $\L$ are required. We can compute the same
subgraphs in parallel, as in the Cholesky decomposition. This time,
however, we first compute the values of the submatrix connected to the
separating set, and referred to as $\Q_S$ in
Fig.~\ref{fig:matrix_perm}. Once we have recursively computed all
necessary inverse elements belonging to the indices of $\Q_S$, we can
determine the inverse elements of $\Sigma$ with indices belonging to
$\Q_A$ and $\Q_B$ in parallel. This is possible since $\Q_A$ and
$\Q_B$ are only connected through $\Q_S$ and these entries have
already been computed at this point. Since for larger matrices, each
submatrix $\Q_A, \Q_B$ was again permuted following the same strategy,
we can recursively traverse the matrix, in the opposite direction as
in the Cholesky decomposition, computing the required entries of
$\Sigma$ in many parallel regions~\cite{bollhofer2020state}.

PARDISO uses OpenMP to carry out the simultaneous computations. Thus,
we obtain a nested OpenMP structure, as we have multiple matrix
factorizations or inversions carried out in parallel, see
Sec.~\ref{subSecParGradient} and \ref{subSecParLineSearch}.
Additionally to the parallelism within PARDISO, OpenMP is used on the
second level to parallelize matrix operations e.g. during the
computation of matrix-matrix or matrix-vector products and while
assembling matrices.

% In the following we will introduce different applications to see the
% improved performance of the previously discussed concepts in action.

% \section{Applications}\label{sec:applications}

\section{Results \& Benchmarks}\label{sec:results}

In this section we present performance results of the newly introduced
parallelization strategies for various applications. They are based on
previously published models that make use of the R-INLA package. We
will not go into the details of the individual application as all the
details can be found in the original publications, and instead focus
on the computational aspects involved. All numerical experiments for
Case Study I \& II were performed on a single node machine with 755 GB
of main memory and 26 dual-socket Intel(R) Xeon(R) Gold 6230R CPU @
2.10GHz, totaling 52 cores. The large number of cores allows us to
demonstrate the full performance gains that can be obtained through
parallelism. All numerical experiments for Case Study III were
performed on an Apple M1 Mac mini with 16 GB of memory, 4 performance
and 4 efficiency cores.
We chose this machine to illustrate that we also obtain performance gains through parallelism on regular desktop computers, laptops and notebooks, although of course to a smaller extent than compared to a larger computer architecture.\\
\smallskip

\textbf{Case Study I : Joint survival modeling of randomized clinical trial.} \\
Rustand et al. recently presented in~\cite{art698}, with R package
\textbf{INLAjoint}\footnote{\url{https://github.com/DenisRustand/INLAjoint}},
a joint survival model consisting of multivariate longitudinal markers
paired with competing risks of events. The different submodels are
linked to each other through shared or correlated random effects.
% , which poses a computational challenge due
The authors consider a randomized placebo controlled trial for the
treatment of the rare autoimmune disease primary biliary cholangitis
(PBC) which affects the liver.
% overview model
Their model has a particularly large number of hyperparamters,
$d(\th) = 50$, due to the 7 different correlated likelihoods.
% used empirical Bayes strategy. provide a unified framework to
% describe

\begin{table}[h!]
    \centering
    \begin{tabular}{l|l|l|l}
      & d($\th$)              & \# lat. par & \# of obs.  \\ \hline \hline
      CS I    & 50 &   51 290  &   27 330   \\                
    \end{tabular}
    \vspace{1mm} \captionsetup{justification=centering}
    \caption{Overview parameters Case Study I}
    \vspace*{-\baselineskip}
\end{table}
The model setup allows them to detect more complex dependencies
between the various biomarkers and the different competing risks, that
go unseen in simpler approaches. \smallskip

% , consisting of submodels that are linked through shared or
% correlated random effects,
\textbf{Case Study II : Cortical surface modeling of human brain activation.} \\
% \textcolor{red}{Add package \& version.}
Spencer et al. show in \cite{spencer2022spatial} that functional brain
responses can be reliably estimated using cortical surface-based
spatial Bayesian generalized linear models (GLMs).
% something fMRI
Functional magnetic resonance imaging (fMRI) data from individual
subjects is used to identify areas of significant activation during a
task or stimulus.
% something about the model
The authors use the stochastic partial differential equations approach
\cite{lindgren2011explicit,art691,book126} for manifolds to define a
spatial GMRF field on the surface of the brain employing geodesic
distances. This allows to flexibly encode spatial dependency
structures in the model while maintaining sparsity in the precision
matrices.
% something more what the model does Their model for a individual
% subject with single task repetition takes the following form
Their approach defines a GLM that fits the framework presented in
\ref{subsec:LGM}. The likelihood is assumed to follow a Gaussian
distribution and we therefore do not require an inner optimization
loop for every function evaluation. More details can be found
in~\cite{spencer2022spatial, mejia2020bayesian} which also includes
information about the corresponding R package
\textbf{BayesfMRI}\footnote{\url{https://github.com/mandymejia/BayesfMRI}(ver.0.1.8),
    \url{https://github.com/danieladamspencer/BayesGLM\_Validation}}.
% something about the setup
For this work we fit a single subject, single repetition set up using
fMRI data for 4 tasks. The dimension of the hyperparameter space
d($\th$) = 9, two for the parametrization of the spatial field of each
task and a noise term for the observations. The images are
preprocessed for standardization and the surface of the brain is
discretized using a triangular mesh. We use two different spatial
discretizations that are determined by the resampling size, which
subsequently influence the dimension of the latent parameter space and
the number of considered observations.

\begin{table}[h!]
    \centering
    \begin{tabular}{l|l|l|l|l}
      & d($\th$)     &       res. size       & \# lat. par & \# of obs.  \\ \hline \hline
      med. CSII       & 9 &   \ 5 000 & \ 35 544 &   \ 2 541 396 \\
      large CSII      & 9     & 25 000 &  183 624   &    13 129 116
    \end{tabular}
    \vspace{1mm} \captionsetup{justification=centering}
    \caption{Overview parameters Case Study II}
    \vspace*{-\baselineskip}
\end{table}
\smallskip

\textbf{Case Study III : Spatial variation in Leukemia survival data.} \\
As a final example we consider a study on spatial variation in
Leukemia survival data from~\cite{henderson2002modeling,
    lindgren2011explicit}. The additive linear predictor of the hazard
function includes 5 fixed effects and a spatial GMRF random field,
which induces a three-dimensional hyperparamter space.
% using classical model to make problem a bit larger ...
\begin{table}[h!]
    \centering
    \begin{tabular}{l|l|l|l}
      & d($\th$)              & \# lat. par & \# of obs.  \\ \hline \hline
      CS III  & 3 &   7945     &   6174   \\
    \end{tabular}
    \vspace{1mm} \captionsetup{justification=centering}
    \caption{Overview parameters Case Study III}
    \vspace*{-\baselineskip}
\end{table}
The fitted model provides survival estimates given the covariates and
spatial locations.
\subsection{Leveraging Level 1 Parallelism}

\begin{figure*}[ht!]
    \centering \subfigure[Case Study I]{
        \begin{tikzpicture}[scale=0.68]
            \begin{axis}
                [%legend pos={north west},
                xlabel = {\# threads level 1}, ylabel = {time per fn
                    in sec},
                ymin=0, ymax=1.2, ymajorgrids=true, grid style=dashed,
                xtick pos=left, ytick pos=left, every axis
                plot/.append style={thick}]
                                
                \pgfplotstableread{jointModel/med_results_pcb_l2fixed_PLS_F_fullInt.txt}\loadedtable
                \addplot [mark=square*, myblue] table[col sep=comma,
                x="threadsL1",y="timePerFuncCall"] {\loadedtable};
                % \addplot [mark=square*, myblue] table[col sep=comma,
                % x="threadsL1",y="totalTime"] {\loadedtable};
            \end{axis}
            \begin{axis}
                [axis y line=right, axis line style={-}, axis x
                line=none, ymin=0, ymax=18, ylabel=speedup per fn]
                                
                \pgfplotstableread{jointModel/med_results_pcb_l2fixed_PLS_F_fullInt.txt}\loadedtable
                \addplot [mark=square*, myblue, dashed] table[col
                sep=comma, x="threadsL1",y="speedupPerIter"]
                {\loadedtable};
                % \addplot [mark=square*, myblue, dashed] table[col
                % sep=comma, x="threadsL1",y="speedup"]
                % {\loadedtable};
            \end{axis}
        \end{tikzpicture}%
    }%
    % FINAL RESULTS
    \hspace{5pt} \subfigure[large Case Study II]{
        \begin{tikzpicture}[scale=0.68]
            \begin{axis}
                [%legend pos={north west},
                xlabel = {\# threads level 1}, ylabel = {time per fn
                    in sec},
                ymin=0, ymax=19, ymajorgrids=true, grid style=dashed,
                xtick pos=left, ytick pos=left, every axis
                plot/.append style={thick}]
                                
                \pgfplotstableread{fMRIModel/med_results_rs25000_l2fixed_final.txt}\loadedtable
                \addplot [mark=square*, myblue] table[col sep=comma,
                x="threadsL1",y="timePerFuncCall"] {\loadedtable};
                % \addplot [mark=square*, myblue] table[col sep=comma,
                % x="threadsL1",y="totalTime"] {\loadedtable};
                                
                % \legend{Pardiso, GPU},
                                
            \end{axis}
                        
                        \begin{axis}
                            [axis y line=right, axis line style={-},
                            axis x line=none, ymin=0, ymax=7.9,
                            ylabel=speedup per fn]
                                
                            \pgfplotstableread{fMRIModel/med_results_rs25000_l2fixed_final.txt}\loadedtable
                            \addplot [mark=square*, myblue, dashed]
                            table[col sep=comma,
                            x="threadsL1",y="speedupPerIter"]
                            {\loadedtable};
                            % \addplot [mark=square*, myblue, dashed]
                            % table[col sep=comma,
                            % x="threadsL1",y="speedup"]
                            % {\loadedtable};
                                
                            % \legend{Pardiso, GPU},
                                
                        \end{axis}
                    \end{tikzpicture}%
                }%end subfigure
        \caption{
                %\textcolor{red}{This is using full integration scheme aka simplified Laplace approximation.}
                The solid line shows the total runtime in seconds divided by the number of function evaluations for different numbers of threads on level 1 with the number of level 2 threads fixed to one and without the parallel linesearch enabled. The dashed line shows the relative speedup over the single-threaded version. 
                %The left panel shows results for Case Study I. The right panel shows results for the large Case Study II.
            }
            \label{fig:scaling_L1}
        \end{figure*}
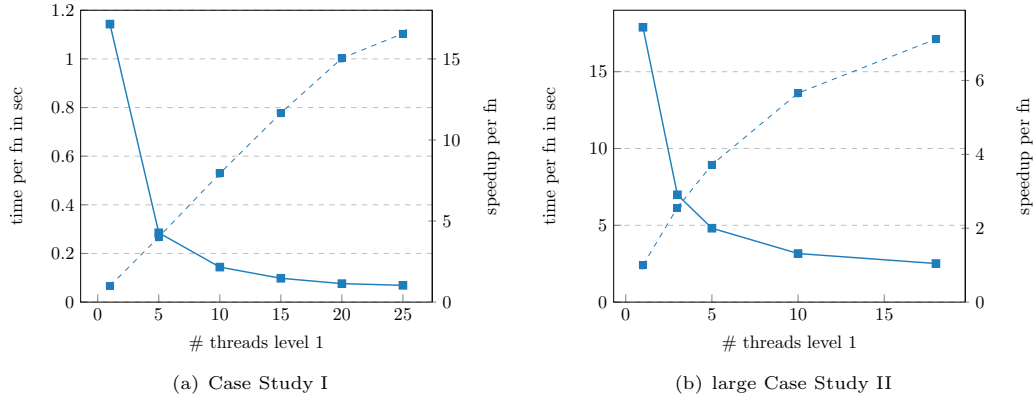

        The level 1 parallelism enables simultaneous function
        evaluations as described in Sec.~\ref{subSecParGradient} which
        are relevant for the gradient computation, the final Hessian
        approximation during the optimization phase as well as the
        partial inversion scheme at the mode.
        % and during the parallel linesearch which will be discussed
        % separately in the next section.
        The speedup that can theoretically be obtained through
        parallelization depends on the number of hyperparamters. If a
        forward difference scheme is employed a maximum speedup of
        dim($\th$)+1 is possible during the gradient computation, for
        a central difference scheme it is
        $2 \- \cdot \- \text{dim}(\th)$. If not specified otherwise
        R-INLA employs a mix of forward and central difference
        approximations. The theoretical speedup during the Hessian
        approximation and the partial inversion scheme depends on the
        number of integration points which in turn depends on the
        chosen integration scheme as well as the dimension of $\th$.
        While these are the computationally most costly operations,
        there are other non-parallelizable steps required.
        Additionally, setting up a parallel OpenMP environment always
        creates overhead. Hence, the theoretical speedup can almost
        never be attained. For relatively small applications the
        observed speedup through parallelism is usually not as
        significant as for larger problems, because the sequential
        parts and the induced overhead make up a larger part of the
        total time. On the other hand, larger applications using a
        very large number of threads might not exhibit further speed
        up after a certain thread count as the memory bandwidth can
        become the limiting factor. We will be able to observe both
        effects when looking at performance results of parallelizing
        level 1. In Fig.~\ref{fig:scaling_L1} we provide scaling
        results with varying number of threads on level 1 and a fixed
        number of threads on level 2 for Case Studies I \& II. Instead
        of showing the total runtime for each case, we divide it by
        the total number of function evaluations of $f$, as the number
        of $f$ evaluations can vary a bit depending on randomly set
        initial values and numerical imprecision due to different
        round-off errors.
        % FINAL RESULTS
        For Case Study I we observe an almost ideal reduction in time
        per $f$ evaluation until 20 threads, yielding an impressive
        speed of a factor of 15 at 20 threads compared to the
        single-threaded version. When employing more than 20 threads
        the parallel efficiency starts to reduce. Case Study II has
        less hyperparameters, hence the maximum attainable speedup on
        level 1 is lower, nevertheless providing significant
        improvements for up to 10 threads with a speedup of almost 6
        compared to the single-threaded version.

        % Uses Case Study III final results!
        We also analyse the effects of level 1 parallelism for Case
        Study III. This model is of much smaller dimension than the
        first two and hence a larger part of the overall runtime is
        dedicated to sequential operations like setting up or the
        initializations of the parallel regions. Nevertheless we
        observe a speedup of a factor of 2 when using 4 instead of 1
        thread.

\begin{table}[h!]
    \centering
    \begin{tabular}{l|l|l|l|l}
      \# threads level 1& 1   &       2       & 3 & 4  \\ \hline \hline
      time per fn in sec      & 0.031 &  0.023         &   0.017 &  0.016 \\  \hline
      speedup        & 1 & 1.4        & 1.8 & 2 \\           
    \end{tabular}
    \vspace{1mm} \captionsetup{justification=centering}
    \caption{Effects of using different numbers of threads on level 1
        for Case Study III.}
    \vspace*{-\baselineskip}
\end{table}

\subsection{Leveraging Parallel Line Search}

% FINAL RESULTS even though they are not as nice ...
\begin{figure}[h!]
    \centering
    \begin{tikzpicture}[scale=0.68]
        \begin{axis}
            [legend pos={north east}, xlabel = {\# of OMP threads
                level 1}, ylabel = {runtime in sec}, ymin=0, ymax=850,
            ymajorgrids=true, grid style=dashed, xtick pos=left, ytick
            pos=left, every axis plot/.append style={thick}]
                        
            \pgfplotstableread{jointModel/med_results_pcb_l2fixed_PLS_F_final.txt}\loadedtable
            \addplot [mark=square*, myblue]
            table[x="threadsL1",y="totalTime"] {\loadedtable};
            % \addplot [mark=square*, myblue] table[col sep=comma,
            % x="threadsL1",y="timePerFuncCall"] {\loadedtable};
                        
            \pgfplotstableread{jointModel/med_results_pcb_l2fixed_PLS_T_final.txt}\loadedtable
            \addplot [mark=square*, myred]
            table[x="threadsL1",y="totalTime"] {\loadedtable};
            \legend{serial linsearch, parallel linsearch},
                        
        \end{axis}
    \end{tikzpicture}%
    \caption{Runtime for Case Study I, with and without parallel line
        search over varying numbers of threads on level 1. The number
        of threads on level 2 is fixed to 1. The posterior marginals
        of the latent parameters are approximated using the empirical
        Bayes integration strategy.}
    \label{fig:parallel_linesearch_results}
\end{figure}
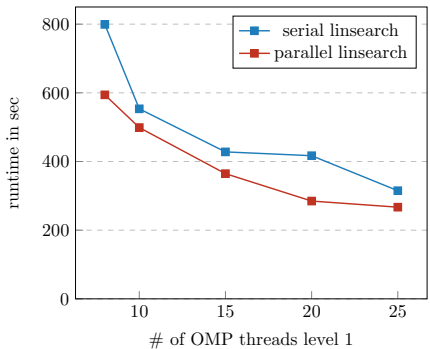

The accuracy of robust regression clearly depends on the number of
available regressors, which in turn is dependent on the number of
available threads. In our experience making use of the parallel line
search implementation only becomes advantageous if 8 threads or more
are in use on level 1. Then there are sufficiently many evaluation
points in the search interval $I$ to adequately represent the original
function. The parallel linesearch becomes especially relevant for
models with non-Gaussian likelihoods as in this case the inaccuracies
in the function evaluations increase due to the arising inner
iteration. In Fig.~\ref{fig:parallel_linesearch_results}, we show the
total runtime of Case Study I using varying numbers of threads on
level 1 with and without enabling the parallel line search. Since this
model has a large number of hyperparameters the computation of the
posterior marginals of the latent parameters using the simplified
Laplace approximation strategy would be the dominating the overall
compute time. In this section we, therefore, used the computationally
cheaper empirical Bayes' approximation, to be able to put the emphasis
on the optimization phase of the algorithm, where the parallel line
search feature is relevant.
% FINAL RESULTS
We can see that it lowers the overall time to solution without
requiring more threads. It additionally exhibits a more stable
behavior when it comes to reduction in runtime over an increasing
thread count.

\subsection{Leveraging Level 2 Parallelism}

The level 2 parallelism occurs within each call of the PARDISO library
as well as other matrix operations like matrix-matrix multiplications,
following the concepts described in Sec.~\ref{subSecParPARDISO}. The
parallelization on level 2 only starts to make a noticeable difference
for larger latent parameter spaces. The single-threaded version of the
PARDISO library is already very efficient for small to moderate sized
problems. The full effects of the parallelization start becoming
visible for the latent parameter spaces of dimensions in the range of
$10^4$ and larger, especially for models with a three-dimensional
structure.

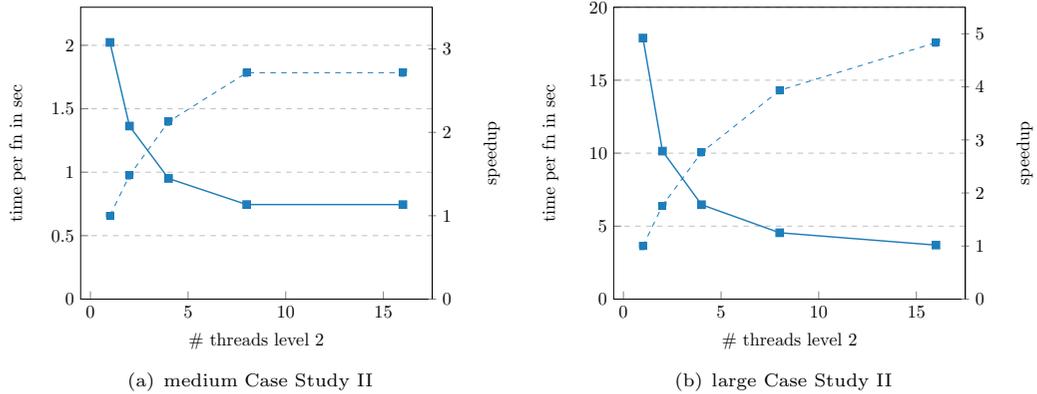
\begin{figure*}[h!]
    \centering \subfigure[medium Case Study II]{
        \begin{tikzpicture}[scale=0.68]
            \begin{axis}
                [%legend pos={north west},
                xlabel = {\# threads level 2}, ylabel = {time per fn
                    in sec}, ymin=0, ymax=2.3, ymajorgrids=true, grid
                style=dashed, xtick pos=left, ytick pos=left, every
                axis plot/.append style={thick}]
                                
                \pgfplotstableread{fMRIModel/med_results_rs5000_l1fixed_final.txt}\loadedtable
                \addplot [mark=square*, myblue] table[col sep=comma,
                x="threadsL2",y="timePerFuncCall"] {\loadedtable};
                % \addplot [mark=square*, myblue] table[
                % x="threadsL2",y="totalTime"] {\loadedtable};
            \end{axis}
                        
                        \begin{axis}
                            [axis y line=right, axis line style={-},
                            axis x line=none, ymin=0, ymax=3.5,
                            ylabel=speedup]
                                
                            \pgfplotstableread{fMRIModel/med_results_rs5000_l1fixed_final.txt}\loadedtable
                            % \addplot [mark=square*, myblue]
                            % table[col sep=comma,
                            % x="threadsL2",y="timePerFuncCall"]
                            % {\loadedtable};
                            \addplot [mark=square*, myblue, dashed]
                            table[x="threadsL2",y="speedupPerIter"]
                            {\loadedtable};
                        \end{axis}
                        
                \end{tikzpicture}%
            }%end subfigure
            % FINAL RESULTS
            \hspace{5pt} \subfigure[large Case Study II]{
                \begin{tikzpicture}[scale=0.68]
                    \begin{axis}
                        [%legend pos={north west},
                        xlabel = {\# threads level 2}, ylabel = {time
                            per fn in sec}, ymin=0, ymax=20,
                        ymajorgrids=true, grid style=dashed, xtick
                        pos=left, ytick pos=left, every axis
                        plot/.append style={thick}]
                                
                        \pgfplotstableread{fMRIModel/med_results_rs25000_l1fixed_final.txt}\loadedtable
                        \addplot [mark=square*, myblue] table[col
                        sep=comma, x="threadsL2",y="timePerFuncCall"]
                        {\loadedtable};
                        % \addplot [mark=square*, myblue] table[col
                        % sep=comma, x="threadsL2",y="totalTime"]
                        % {\loadedtable};
                                
                    \end{axis}
                        
                        \begin{axis}
                            [axis y line=right, axis line style={-},
                            axis x line=none, ymin=0, ymax=5.5,
                            ylabel=speedup]
                                
                            \pgfplotstableread{fMRIModel/med_results_rs25000_l1fixed_final.txt}\loadedtable
                            % \addplot [mark=square*, myblue]
                            % table[col sep=comma,
                            % x="threadsL2",y="timePerFuncCall"]
                            % {\loadedtable};
                            \addplot [mark=square*, myblue, dashed]
                            table[col sep=comma,
                            x="threadsL2",y="speedupPerIter"]
                            {\loadedtable};
                                
                        \end{axis}
                        
                \end{tikzpicture}%
            } %end subfigure
        \caption{The solid line shows the total runtime in seconds divided by the number of function evaluations for different numbers of threads on level 2 with the number of level 1 threads fixed to one and without the parallel linesearch enabled. The dashed line shows the relative speedup over the single-threaded version. 
        %The left panel shows results for the medium-sized Case Study II. The right panel shows results for the large Case Study II.
        }
        \label{fig:scaling_L2}
    \end{figure*}

    We consider both examples of Case Study II in the performance
    analysis of Fig. ~\ref{fig:scaling_L2}.
    % FINAL RESULTS
    For a fixed number of threads on level 1, we observe a continuous
    speedup for increasing numbers of threads on level 2 up until 8
    threads for the medium-sized Case Study II, leading to a maximum
    speedup of almost 3 over the single-threaded version. In turn the
    larger-sized Case Study II continues to exhibit a speedup also
    beyond 8 threads, showing that larger latent parameter space
    benefit more from level 2 parallelism, leading to a speedup of
    almost 5 at 16 threads over the single-threaded version.
    % measure overall run time -> clearly many non-parallel section,
    % matrix assembly etc.

    \subsection{Leveraging combined Parallelism}

    % ideally also include example that runs more smoothly with
    % parallel line search
    \begin{figure*}[ht!]
        \centering \subfigure[Case Study I]{
            \begin{tikzpicture}[scale=0.63]
                % FINAL RESULTS -- FULL INTEGRATION STRATEGY
                \begin{axis}
                    [ ybar, enlargelimits=0.15, ytick
                    style={draw=none}, % remove ticks y axis
                    yticklabels={,,}, % remove numbers from y axis
                    scaled y ticks=false, % to remove 10^4
                    ybar=-20, % need because of bar width
                    ylabel={runtime in
                        sec}, % the ylabel must precede a # symbol.
                    compat=1.3, % moves y label closer to axis
                    symbolic x coords = {1:1, 5:1, 10:1, 20:1, 30:1,
                        40:1}, xtick={1:1, 5:1, 10:1, 20:1, 30:1,
                        40:1}, xtick pos=left, nodes near
                    coords, % this command is used to mention the y-axis points on the top of the particular bar.
                    nodes near coords align={vertical}, bar width=20,
                    ]
                    % put run times in here
                    % plot rs 30 000, PLS F, different thread counts
                    \addplot[myblue, fill] coordinates {(1:1,24331)};
                    \addplot[myblue, fill] coordinates {(5:1,5539)};
                    \addplot[myblue, fill] coordinates {(10:1,3095)
                        (20:1,1682)}; \addplot[myblue, fill]
                    coordinates {(30:1,1152) (40:1,996)};
                \end{axis}
            \end{tikzpicture}
        }%end subfigure
        % FINAL RESULTS
        \subfigure[medium Case Study II]{
            \begin{tikzpicture}[scale=0.63]
                \begin{axis}
                    [ ybar, enlargelimits=0.15, ytick
                    style={draw=none}, % remove ticks y axis
                    yticklabels={,,}, % remove numbers from y axis
                    ybar=-20, % need because of bar width
                    ylabel={runtime in
                        sec}, % the ylabel must precede a # symbol.
                    compat=1.3, % moves y label closer to axis
                    symbolic x coords = {1:1,3:1, 1:4, 9:1, 5:4, 9:4},
                    xtick={1:1, 3:1, 1:4, 9:1, 5:4, 9:4}, xtick
                    pos=left, nodes near
                    coords, % this command is used to mention the y-axis points on the top of the particular bar.
                    nodes near coords align={vertical}, bar width=20,
                    ]
                    % put run times in here
                    % plot rs 30 000, PLS F, different thread counts
                    \addplot[myblue, fill] coordinates
                    {(1:1,1149)(3:1, 496)}; \addplot[myblue, fill]
                    coordinates {(1:4,561)(9:1,147)}; \addplot[myblue,
                    fill] coordinates {(5:4,84)(9:4,82) };
                                
                \end{axis}
            \end{tikzpicture}
        }% end subfigure
        \subfigure[large Case Study II]{
            \begin{tikzpicture}[scale=0.63]
                % TODO: not yet final results
                \begin{axis}
                    [ ybar, enlargelimits=0.15, ytick
                    style={draw=none}, % remove ticks y axis
                    yticklabels={,,}, % remove numbers from y axis
                    scaled y ticks=false, % to remove 10^4
                    ybar=-20, % need because of bar width
                    ylabel={runtime in
                        sec}, % the ylabel must precede a # symbol.
                    compat=1.3, % moves y label closer to axis
                    symbolic x coords = {1:1, 3:1, 1:8, 10:1, 18:2,
                        10:4}, xtick={1:1, 3:1, 1:8, 10:1,18:2, 10:4},
                    xtick pos=left, nodes near
                    coords, % this command is used to mention the y-axis points on the top of the particular bar.
                    nodes near coords align={vertical}, bar width=20,
                    ]
                    % put run times in here
                    % plot rs 30 000, PLS F, different thread counts
                    \addplot[myblue, fill] coordinates
                    {(1:1,6763)(3:1, 2906)}; \addplot[myblue, fill]
                    coordinates {(1:8,1719)(10:1,1195)};
                    \addplot[myblue, fill] coordinates {(18:2,868)
                        (10:4,611)};
                \end{axis}
            \end{tikzpicture}
        }
        \caption{Runtime in seconds for different thread
            configurations. The first number represents the threads on
            level 1, the second number the threads on level 2. The
            total number of threads is equal to (\# threads level 1)
            $\cdot$ (\# threads level 2).}
        \label{fig:overall_results}
    \end{figure*}
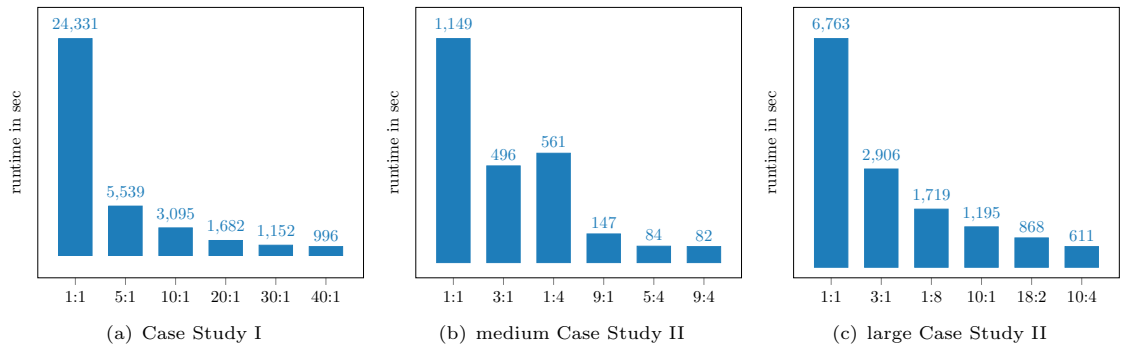

    In this section we want to present results for combined
    parallelization strategies and discuss how to choose a favorable
    set up. The ideal approach clearly depends on the available
    infrastructure and the problem at hand. In almost all cases the
    number of available cores will be limited. Often this limit will
    easily be reached as the total number of threads is equal to (\#
    threads L1) $\cdot$ (\# threads L2). This poses the question of
    how best utilize the available resources. We have found that
    hyperthreading often does not significantly increase the
    performance (or is even counter productive) and therefore only
    consider thread counts that are within the number of physical
    cores. In the previous two sections we have seen that smaller
    problems benefit less from larger thread counts as the overhead of
    thread initialization can overshadow the gain. We can see from
    both Fig.~\ref{fig:scaling_L1}~\&~\ref{fig:scaling_L2} that the
    parallel efficiency is highest for smaller thread counts on both
    levels. This implies that its often favorable to distribute the
    available number of threads among both levels. Additionally the
    structure of the model should be taking into account. If the
    number of hyperparamters is large it is advantageous to focus more
    on level 1 parallelism whereas a very high-dimensional latent
    parameter space will benefit more from level 2 parallelism. This
    behavior can be observed empirically in Fig.
    \ref{fig:overall_results}, where we present the runtime of the
    INLA algorithm in seconds for different thread configurations.
    % and compare to the previous numerical solver TAUCS. For Case
    % Study III we have a total runtime of 11 seconds when using the
    % single-threaded version with TAUCS as a numerical solver, when
    % PARDISO is used instead the total runtime drops to 7 seconds.
    % For increasingly larger problems the difference in runtime
    % between the two linear solvers expands as the numerical core
    % operations make up increasingly larger parts of the total
    % runtime. Looking at Figure~\ref{fig:overall_results} one can
    % observe that PARDISO exhibits better scaling for increasing
    % matrix sizes.
    It becomes clear that the introduction of parallelism leads to a
    tremendous reduction in runtime. For Case Study I, we observe
    speedups of a factor 20 and more over the single-threaded
    % PARDISO
    version. For the medium-sized Case Study II we observe a speedup
    of a factor up to 15. While for the large Case Study II we observe
    a speedup of more than 10 times compared to the single-threaded
    version.
    % With speedups of up to 40 times over previous implementations
    For sufficiently large multi-core architectures, the newly
    introduced updates open the door to previously unfeasible modeling
    scales, and drastically reduce waiting times for users.
    % to a new era of fast approximate Bayesian inference of
    % unprecedented scale

    \section{Discussion}
    \label{sec:discussion}

    In this paper we presented novel parallelization strategies for
    the INLA methodology. While INLA has always been using optimized
    sparse linear algebra operations to be computationally efficient,
    we have taken this a step further by introducing parallelism using
    OpenMP. This allows for the simultaneous execution of
    computationally expensive tasks which are distributed over a
    specified number of cores. Our approach contains two layers of
    parallelism. On the first layer we have parallelized independent
    function evaluations during INLA's optimization phase to determine
    and explore the posterior mode of the hyperparameters.
    % The function evaluations are necessary for the approximation of
    % the gradient to determine the direction of the next optimization
    % step.
    Each evaluation requires the Cholesky factorization, a
    computationally expensive task, of at least one sparse precision
    matrix whose size is tied to the potentially very high-dimensional
    latent parameter space of the model. We have additionally
    introduced a parallel line search routine using robust regression
    to parallelize and stabilize the search for each next step size,
    thus being able to lower the overall number of required
    iterations. We were also able to parallelize large parts of the
    computation of the posterior marginal variances of the latent
    parameters which require a computationally expensive partial
    matrix inversion. The state-of-the-art sparse solver PARDISO was
    included in INLA to handle the most expensive linear algebra
    operations and makes up the second layer of parallelism. PARDISO
    makes use of intelligent matrix reordering techniques that allow
    large parts of the otherwise sequential computations to be
    parallelized. Thus, PARDISO internally employs OpenMP for
    simultaneous operations within a Cholesky factorization, the
    solution to a linear system of equations or a partial matrix
    inversion.
    % INLA is a framework for performing Bayesian inference relying on
    % a nested approximation strategy and sparse GMRF representations
    % of the latent parameter space.
    We empirically demonstrated the performance and scalability of the
    parallelization strategies on three case studies. Each of them
    uses a different type of model, hence posing different
    computational challenges.
    % TODO: double check with final numbers
    For all larger applications we observed speeds up of a factor of
    10-25 times compared to the single-threaded version.
    % and speedups of 30-40 times compared to the previous
    % implementation.
    Hence, lowering runtimes for some models from almost seven hours
    to less than 30 minutes or from more than two hours to just over
    ten minutes. For smaller applications we still observe speedups
    but to a smaller extend. Given that larger multi-core computer
    architectures are becoming more and more accessible, these
    computational advancements allow users to perform inference using
    more complex models at higher resolution in shorter runtimes with
    INLA.
    % As we will show, these advancements allow users to compute
    % models of previously infeasible latent parameter dimensions,

    \backmatter

\begin{comment}
    \bmhead{Supplementary information}

    \textcolor{red}{Use Case Study III (Leukemia example) to
        demonstrate how to set threads?}

    \textcolor{red}{@haavard: should we put something here? Or make a
        repository with the code somewhere? }
    % If your article has accompanying supplementary file/s please
    % state so here.
\end{comment}

\bmhead{Acknowledgments} We would like to thank Prof. D. Bolin and Dr.
D. Rustand for their support with the different case studies. The work
of L. Gaedke-Merzh\"{a}user has been supported by the SNF SINERGIA
project number CRSII5\_189942.
\bibliographystyle{chicago}
\bibliography{sn-bibliography,mybib}% common bib file

\begin{thebibliography}{}

\bibitem[\protect\citeauthoryear{Arisido, Gaetan, Zanchettin, and
  Rubino}{Arisido et~al.}{2017}]{arisido2017bayesian}
Arisido, M.~W., C.~Gaetan, D.~Zanchettin, and A.~Rubino (2017).
\newblock A {B}ayesian hierarchical approach for spatial analysis of climate
  model bias in multi-model ensembles.
\newblock {\em Stochastic Environmental Research and Risk Assessment\/}~{\em
  31\/}(10), 2645--2657.
\newblock \url{https://doi.org/10.1007/s00477-017-1383-2}.

\bibitem[\protect\citeauthoryear{Ascher and Greif}{Ascher and
  Greif}{2011}]{ascher2011first}
Ascher, U.~M. and C.~Greif (2011).
\newblock {\em A first course on numerical methods}.
\newblock SIAM.
\newblock \url{https://doi.org/10.1137/9780898719987}.

\bibitem[\protect\citeauthoryear{Atkinson, Riani, and Riani}{Atkinson
  et~al.}{2000}]{atkinson2000robust}
Atkinson, A.~C., M.~Riani, and M.~Riani (2000).
\newblock {\em Robust diagnostic regression analysis}, Volume~2.
\newblock Springer.
\newblock \url{https://doi.org/10.1007/978-1-4612-1160-0}.

\bibitem[\protect\citeauthoryear{Bakka, Rue, Fuglstad, Riebler, Bolin, Illian,
  Krainski, Simpson, and Lindgren}{Bakka et~al.}{2018}]{art643}
Bakka, H., H.~Rue, G.~A. Fuglstad, A.~Riebler, D.~Bolin, J.~Illian,
  E.~Krainski, D.~Simpson, and F.~Lindgren (2018).
\newblock Spatial modelling with {R-INLA}: {A} review.
\newblock {\em WIREs Computational Statistics\/}~{\em 10:e1443\/}(6).
\newblock (Invited extended review), \url{10.1002/wics.1443}.

\bibitem[\protect\citeauthoryear{Batomen, Irving, Carabali, Carvalho, Ruggiero,
  and Brown}{Batomen et~al.}{2020}]{batomen2020vulnerable}
Batomen, B., H.~Irving, M.~Carabali, M.~S. Carvalho, E.~D. Ruggiero, and
  P.~Brown (2020).
\newblock Vulnerable road-user deaths in {B}razil: a {B}ayesian hierarchical
  model for spatial-temporal analysis.
\newblock {\em International journal of injury control and safety promotion\/},
  1--9.
\newblock \url{https://doi.org/10.1080/17457300.2020.1818788}.

\bibitem[\protect\citeauthoryear{Baydin, Pearlmutter, Radul, and
  Siskind}{Baydin et~al.}{2018}]{baydin2018automatic}
Baydin, A.~G., B.~A. Pearlmutter, A.~A. Radul, and J.~M. Siskind (2018).
\newblock Automatic differentiation in machine learning: a survey.
\newblock {\em Journal of Marchine Learning Research\/}~{\em 18}, 1--43.
\newblock \url{https://dl.acm.org/doi/abs/10.5555/3122009.3242010}.

\bibitem[\protect\citeauthoryear{Bhatt, Weiss, Cameron, Bisanzio, Mappin,
  Dalrymple, Battle, Moyes, Henry, Eckhoff, et~al.}{Bhatt
  et~al.}{2015}]{bhatt2015effect}
Bhatt, S., D.~Weiss, E.~Cameron, D.~Bisanzio, B.~Mappin, U.~Dalrymple,
  K.~Battle, C.~Moyes, A.~Henry, P.~Eckhoff, et~al. (2015).
\newblock The effect of malaria control on {P}lasmodium falciparum in {A}frica
  between 2000 and 2015.
\newblock {\em Nature\/}~{\em 526\/}(7572), 207--211.
\newblock \url{https://doi.org/10.1038/nature15535}.

\bibitem[\protect\citeauthoryear{Bichot and Siarry}{Bichot and
  Siarry}{2013}]{bichot2013graph}
Bichot, C.-E. and P.~Siarry (2013).
\newblock {\em Graph partitioning}.
\newblock John Wiley \& Sons.
\newblock \url{https://doi.org/10.1007/978-3-319-63962-8_312-1}.

\bibitem[\protect\citeauthoryear{Bollh{\"o}fer, Schenk, Janalik, Hamm, and
  Gullapalli}{Bollh{\"o}fer et~al.}{2020}]{bollhofer2020state}
Bollh{\"o}fer, M., O.~Schenk, R.~Janalik, S.~Hamm, and K.~Gullapalli (2020).
\newblock State-of-the-art sparse direct solvers.
\newblock In {\em Parallel Algorithms in Computational Science and
  Engineering}, pp.\  3--33. Springer.
\newblock \url{https://doi.org/10.1007/978-3-030-43736-7\_1}.

\bibitem[\protect\citeauthoryear{Coll, Pennino, Steenbeek, Sol{\'e}, and
  Bellido}{Coll et~al.}{2019}]{coll2019}
Coll, M., M.~G. Pennino, J.~Steenbeek, J.~Sol{\'e}, and J.~M. Bellido (2019).
\newblock Predicting marine species distributions: complementarity of food-web
  and bayesian hierarchical modelling approaches.
\newblock {\em Ecological Modelling\/}~{\em 405}, 86--101.
\newblock \url{https://doi.org/10.1016/j.ecolmodel.2019.05.005}.

\bibitem[\protect\citeauthoryear{Congdon}{Congdon}{2014}]{congdon2014applied}
Congdon, P. (2014).
\newblock {\em Applied {B}ayesian modelling}, Volume 595.
\newblock John Wiley \& Sons.
\newblock
  \href{https://doi.org/10.1002/9781118895047}{{10.1002/9781118895047}}.

\bibitem[\protect\citeauthoryear{Davis}{Davis}{2006}]{davis2006direct}
Davis, T.~A. (2006).
\newblock {\em Direct methods for sparse linear systems}.
\newblock SIAM.
\newblock \url{https://doi.org/10.1137/1.9780898718881}.

\bibitem[\protect\citeauthoryear{de~Rivera, Blangiardo, L{\'o}pez-Qu{\'\i}lez,
  and Mart{\'\i}n-Sanz}{de~Rivera et~al.}{2019}]{de2019species}
de~Rivera, O.~R., M.~Blangiardo, A.~L{\'o}pez-Qu{\'\i}lez, and
  I.~Mart{\'\i}n-Sanz (2019).
\newblock Species distribution modelling through {B}ayesian hierarchical
  approach.
\newblock {\em Theoretical Ecology\/}~{\em 12\/}(1), 49--59.
\newblock \url{https://doi.org/10.1007/s12080-018-0387-y}.

\bibitem[\protect\citeauthoryear{Demmel}{Demmel}{1997}]{doi:10.1137/1.9781611971446}
Demmel, J.~W. (1997).
\newblock {\em Applied Numerical Linear Algebra}.
\newblock Society for Industrial and Applied Mathematics.
\newblock \url{https://epubs.siam.org/doi/pdf/10.1137/1.9781611971446}.

\bibitem[\protect\citeauthoryear{Diaz, Pophale, Hernandez, Bernholdt, and
  Chandrasekaran}{Diaz et~al.}{2018}]{OpenMP}
Diaz, J.~M., S.~Pophale, O.~Hernandez, D.~E. Bernholdt, and S.~Chandrasekaran
  (2018).
\newblock Openmp 4.5 validation and verification suite for device offload.
\newblock In B.~R. de~Supinski, P.~Valero-Lara, X.~Martorell, S.~Mateo~Bellido,
  and J.~Labarta (Eds.), {\em Evolving OpenMP for Evolving Architectures},
  Cham, pp.\  82--95. Springer International Publishing.
\newblock \url{https://www.openmp.org}.

\bibitem[\protect\citeauthoryear{Fattah, Niekerk, and Rue}{Fattah
  et~al.}{2022}]{fattah2022smart}
Fattah, E.~A., J.~V. Niekerk, and H.~Rue (2022).
\newblock Smart gradient - an adaptive technique for improving gradient
  estimation.
\newblock {\em Foundations of Data Science\/}~{\em 4\/}(1), 123--136.
\newblock \url{https://www.aimsciences.org/article/doi/10.3934/fods.2021037}.

\bibitem[\protect\citeauthoryear{George}{George}{1973}]{george1973nested}
George, A. (1973).
\newblock Nested dissection of a regular finite element mesh.
\newblock {\em SIAM Journal on Numerical Analysis\/}~{\em 10\/}(2), 345--363.
\newblock \url{https://doi.org/10.1137/0710032}.

\bibitem[\protect\citeauthoryear{George and Liu}{George and
  Liu}{1989}]{george1989evolution}
George, A. and J.~W. Liu (1989).
\newblock The evolution of the minimum degree ordering algorithm.
\newblock {\em Siam review\/}~{\em 31\/}(1), 1--19.
\newblock \url{https://doi.org/10.1137/1031001}.

\bibitem[\protect\citeauthoryear{Heath, Ng, and Peyton}{Heath
  et~al.}{1991}]{heath1991parallel}
Heath, M.~T., E.~Ng, and B.~W. Peyton (1991).
\newblock Parallel algorithms for sparse linear systems.
\newblock {\em SIAM review\/}~{\em 33\/}(3), 420--460.
\newblock \url{https://doi.org/10.1137/1033099}.

\bibitem[\protect\citeauthoryear{Henderson, Shimakura, and Gorst}{Henderson
  et~al.}{2002}]{henderson2002modeling}
Henderson, R., S.~Shimakura, and D.~Gorst (2002).
\newblock Modeling spatial variation in leukemia survival data.
\newblock {\em Journal of the American Statistical Association\/}~{\em
  97\/}(460), 965--972.
\newblock \url{https://doi.org/10.1198/016214502388618753}.

\bibitem[\protect\citeauthoryear{Isaac, Jarzyna, Keil, Dambly, Boersch-Supan,
  Browning, Freeman, Golding, Guillera-Arroita, Henrys, et~al.}{Isaac
  et~al.}{2020}]{isaac2020}
Isaac, N.~J., M.~A. Jarzyna, P.~Keil, L.~I. Dambly, P.~H. Boersch-Supan,
  E.~Browning, S.~N. Freeman, N.~Golding, G.~Guillera-Arroita, P.~A. Henrys,
  et~al. (2020).
\newblock Data integration for large-scale models of species distributions.
\newblock {\em Trends in ecology \& evolution\/}~{\em 35\/}(1), 56--67.
\newblock \url{https://doi.org/10.1016/j.tree.2019.08.006}.

\bibitem[\protect\citeauthoryear{Karypis and Kumar}{Karypis and
  Kumar}{1998}]{karypis1998fast}
Karypis, G. and V.~Kumar (1998).
\newblock A fast and high quality multilevel scheme for partitioning irregular
  graphs.
\newblock {\em SIAM Journal on scientific Computing\/}~{\em 20\/}(1), 359--392.
\newblock \url{https://dl.acm.org/doi/10.5555/305219.305248}.

\bibitem[\protect\citeauthoryear{Konstantinoudis, Padellini, Bennett, Davies,
  Ezzati, and Blangiardo}{Konstantinoudis
  et~al.}{2021}]{konstantinoudis2021long}
Konstantinoudis, G., T.~Padellini, J.~Bennett, B.~Davies, M.~Ezzati, and
  M.~Blangiardo (2021).
\newblock Long-term exposure to air-pollution and covid-19 mortality in
  {E}ngland: a hierarchical spatial analysis.
\newblock {\em Environment international\/}~{\em 146}, 106316.
\newblock \url{https://doi.org/10.1016/j.envint.2020.106316}.

\bibitem[\protect\citeauthoryear{Kontis, Bennett, Rashid, Parks,
  Pearson-Stuttard, Guillot, Asaria, Zhou, Battaglini, Corsetti, et~al.}{Kontis
  et~al.}{2020}]{kontis2020}
Kontis, V., J.~E. Bennett, T.~Rashid, R.~M. Parks, J.~Pearson-Stuttard,
  M.~Guillot, P.~Asaria, B.~Zhou, M.~Battaglini, G.~Corsetti, et~al. (2020).
\newblock Magnitude, demographics and dynamics of the effect of the first wave
  of the covid-19 pandemic on all-cause mortality in 21 industrialized
  countries.
\newblock {\em Nature medicine\/}~{\em 26\/}(12), 1919--1928.
\newblock \url{https://www.nature.com/articles/s41591-020-1112-0}.

\bibitem[\protect\citeauthoryear{Krainski, G{\'o}mez-Rubio, Bakka, Lenzi,
  Castro-Camilio, Simpson, Lindgren, and Rue}{Krainski et~al.}{2018}]{book126}
Krainski, E.~T., V.~G{\'o}mez-Rubio, H.~Bakka, A.~Lenzi, D.~Castro-Camilio,
  D.~Simpson, F.~Lindgren, and H.~Rue (2018, December).
\newblock {\em Advanced Spatial Modeling with Stochastic Partial Differential
  Equations using {R} and {INLA}}.
\newblock CRC press.
\newblock Github version \url{www.r-inla.org/spde-book}.

\bibitem[\protect\citeauthoryear{LeVeque}{LeVeque}{2007}]{leveque2007finite}
LeVeque, R.~J. (2007).
\newblock {\em Finite difference methods for ordinary and partial differential
  equations: steady-state and time-dependent problems}.
\newblock SIAM.
\newblock \url{https://doi.org/10.1137/1.9780898717839}.

\bibitem[\protect\citeauthoryear{Li, Ahmed, Klimeck, and Darve}{Li
  et~al.}{2008}]{LI20089408}
Li, S., S.~Ahmed, G.~Klimeck, and E.~Darve (2008).
\newblock Computing entries of the inverse of a sparse matrix using the {FIND}
  algorithm.
\newblock {\em Journal of Computational Physics\/}~{\em 227\/}(22), 9408--9427.
\newblock \url{https://doi.org/10.1016/j.jcp.2008.06.033}.

\bibitem[\protect\citeauthoryear{Lillini, Tittarelli, Bertoldi, Ritchie,
  Katalinic, Pritzkuleit, Launoy, Launay, Guillaume, {\v{Z}}agar,
  et~al.}{Lillini et~al.}{2021}]{lillini2021}
Lillini, R., A.~Tittarelli, M.~Bertoldi, D.~Ritchie, A.~Katalinic,
  R.~Pritzkuleit, G.~Launoy, L.~Launay, E.~Guillaume, T.~{\v{Z}}agar, et~al.
  (2021).
\newblock Water and soil pollution: Ecological environmental study
  methodologies useful for public health projects. a literature review.
\newblock {\em Reviews of Environmental Contamination and Toxicology Volume
  256\/}, 179--214.
\newblock \url{https://doi.org/10.1007/398_2020_58}.

\bibitem[\protect\citeauthoryear{Lindenmayer, Taylor, and
  Blanchard}{Lindenmayer et~al.}{2021}]{lindenmayer2021}
Lindenmayer, D., C.~Taylor, and W.~Blanchard (2021).
\newblock Empirical analyses of the factors influencing fire severity in
  southeastern australia.
\newblock {\em Ecosphere\/}~{\em 12\/}(8), e03721.
\newblock \url{ https://doi.org/10.1002/ecs2.3721}.

\bibitem[\protect\citeauthoryear{Lindgren, Bolin, and Rue}{Lindgren
  et~al.}{2022}]{art691}
Lindgren, F., D.~Bolin, and H.~Rue (2022).
\newblock The {SPDE} approach for {G}aussian and non-{G}aussian fields: 10
  years and still running.
\newblock {\em Spatial Statistics\/}~{\em xx\/}(xx), xx--xx.
\newblock (accepted), \url{https://doi.org/10.1016/j.spasta.2022.100599}.

\bibitem[\protect\citeauthoryear{Lindgren, Rue, and Lindstr{\"o}m}{Lindgren
  et~al.}{2011}]{lindgren2011explicit}
Lindgren, F., H.~Rue, and J.~Lindstr{\"o}m (2011).
\newblock An explicit link between {G}aussian fields and {G}aussian {M}arkov
  random fields: the stochastic partial differential equation approach.
\newblock {\em Journal of the Royal Statistical Society: Series B (Statistical
  Methodology)\/}~{\em 73\/}(4), 423--498.
\newblock
  \href{https://doi.org/10.1111/j.1467-9868.2011.00777.x}{{10.1111/j.1467-9868.2011.00777.x}}.

\bibitem[\protect\citeauthoryear{Lu, Liang, Huang, Qin, Yao, Wang, and Yang}{Lu
  et~al.}{2018}]{lu2018hierarchical}
Lu, N., S.~Liang, G.~Huang, J.~Qin, L.~Yao, D.~Wang, and K.~Yang (2018).
\newblock Hierarchical {B}ayesian space-time estimation of monthly maximum and
  minimum surface air temperature.
\newblock {\em Remote Sensing of Environment\/}~{\em 211}, 48--58.
\newblock
  \href{https://doi.org/10.1016/j.rse.2018.04.006}{{10.1016/j.rse.2018.04.006}}.

\bibitem[\protect\citeauthoryear{Mart{\'\i}nez-Minaya, Cameletti, Conesa, and
  Pennino}{Mart{\'\i}nez-Minaya et~al.}{2018}]{martinez2018}
Mart{\'\i}nez-Minaya, J., M.~Cameletti, D.~Conesa, and M.~G. Pennino (2018).
\newblock Species distribution modeling: a statistical review with focus in
  spatio-temporal issues.
\newblock {\em Stochastic environmental research and risk assessment\/}~{\em
  32\/}(11), 3227--3244.
\newblock \url{https://doi.org/10.1007/s00477-018-1548-7}.

\bibitem[\protect\citeauthoryear{Martins, Simpson, Lindgren, and Rue}{Martins
  et~al.}{2013}]{martins2013bayesian}
Martins, T.~G., D.~Simpson, F.~Lindgren, and H.~Rue (2013).
\newblock Bayesian computing with inla: new features.
\newblock {\em Computational Statistics \& Data Analysis\/}~{\em 67}, 68--83.
\newblock
  \href{https://doi.org/10.1016/j.csda.2013.04.014}{{10.1016/j.csda.2013.04.014}}.

\bibitem[\protect\citeauthoryear{Mejia, Yue, Bolin, Lindgren, and
  Lindquist}{Mejia et~al.}{2020}]{mejia2020bayesian}
Mejia, A.~F., Y.~Yue, D.~Bolin, F.~Lindgren, and M.~A. Lindquist (2020).
\newblock A bayesian general linear modeling approach to cortical surface fmri
  data analysis.
\newblock {\em Journal of the American Statistical Association\/}~{\em
  115\/}(530), 501--520.
\newblock \url{https://doi.org/10.1080/01621459.2019.1611582}.

\bibitem[\protect\citeauthoryear{Mielke, Claassen, Busana, Heskes, Huijbregts,
  Koffijberg, and Schipper}{Mielke et~al.}{2020}]{mielke2020}
Mielke, K.~P., T.~Claassen, M.~Busana, T.~Heskes, M.~A. Huijbregts,
  K.~Koffijberg, and A.~M. Schipper (2020).
\newblock Disentangling drivers of spatial autocorrelation in species
  distribution models.
\newblock {\em Ecography\/}~{\em 43\/}(12), 1741--1751.
\newblock \url{ https://doi.org/10.1111/ecog.05134}.

\bibitem[\protect\citeauthoryear{Nocedal and Wright}{Nocedal and
  Wright}{2006}]{nocedal2006numerical}
Nocedal, J. and S.~Wright (2006).
\newblock {\em Numerical optimization}.
\newblock Springer Science \& Business Media.
\newblock
  \href{https://doi.org/10.1007/978-0-387-40065-5}{{10.1007/978-0-387-40065-5}}.

\bibitem[\protect\citeauthoryear{Opitz}{Opitz}{2017}]{opitz2017latent}
Opitz, T. (2017).
\newblock Latent gaussian modeling and inla: A review with focus on space-time
  applications.
\newblock {\em Journal de la soci{\'e}t{\'e} fran{\c{c}}aise de
  statistique\/}~{\em 158\/}(3), 62--85.
\newblock
  \href{https://hal.archives-ouvertes.fr/hal-01394974/document}{{hal.archives-ouvertes.fr/hal-01394974}}.

\bibitem[\protect\citeauthoryear{Pan and Reif}{Pan and
  Reif}{1985}]{pan1985efficient}
Pan, V. and J.~Reif (1985).
\newblock Efficient parallel solution of linear systems.
\newblock In {\em Proceedings of the seventeenth annual ACM symposium on Theory
  of computing}, pp.\  143--152.
\newblock \url{https://doi.org/10.1145/22145.22161}.

\bibitem[\protect\citeauthoryear{Pimont, Fargeon, Opitz, Ruffault, Barbero,
  Martin-StPaul, Rigolot, Rivi{\`e}re, and Dupuy}{Pimont
  et~al.}{2021}]{pimont2021}
Pimont, F., H.~Fargeon, T.~Opitz, J.~Ruffault, R.~Barbero, N.~Martin-StPaul,
  E.~Rigolot, M.~Rivi{\`e}re, and J.-L. Dupuy (2021).
\newblock Prediction of regional wildfire activity in the probabilistic
  bayesian framework of firelihood.
\newblock {\em Ecological applications\/}~{\em 31\/}(5), e02316.
\newblock \url{https://doi.org/10.1002/eap.2316}.

\bibitem[\protect\citeauthoryear{Pinto, Rousseu, Niklasson, and
  Drobyshev}{Pinto et~al.}{2020}]{pinto2020}
Pinto, G., F.~Rousseu, M.~Niklasson, and I.~Drobyshev (2020).
\newblock Effects of human-related and biotic landscape features on the
  occurrence and size of modern forest fires in sweden.
\newblock {\em Agricultural and Forest Meteorology\/}~{\em 291}, 108084.
\newblock \url{https://doi.org/10.1016/j.agrformet.2020.108084}.

\bibitem[\protect\citeauthoryear{Rousseeuw and Leroy}{Rousseeuw and
  Leroy}{2005}]{rousseeuw2005robust}
Rousseeuw, P.~J. and A.~M. Leroy (2005).
\newblock {\em Robust regression and outlier detection}, Volume 589.
\newblock John wiley \& sons.
\newblock \url{https://doi.org/10.1002/0471725382}.

\bibitem[\protect\citeauthoryear{Rue and Held}{Rue and
  Held}{2005}]{rue2005gaussian}
Rue, H. and L.~Held (2005).
\newblock {\em {G}aussian {M}arkov random fields: theory and applications}.
\newblock CRC press.
\newblock
  \href{https://doi.org/10.1201/9780203492024}{{10.1201/9780203492024}}.

\bibitem[\protect\citeauthoryear{Rue and Martino}{Rue and
  Martino}{2007}]{rue2007approximate}
Rue, H. and S.~Martino (2007).
\newblock Approximate bayesian inference for hierarchical gaussian markov
  random field models.
\newblock {\em Journal of statistical planning and inference\/}~{\em
  137\/}(10), 3177--3192.
\newblock
  \href{https://doi.org/10.1016/j.jspi.2006.07.016}{{10.1016/j.jspi.2006.07.016}}.

\bibitem[\protect\citeauthoryear{Rue, Martino, and Chopin}{Rue
  et~al.}{2009}]{rue2009approximate}
Rue, H., S.~Martino, and N.~Chopin (2009).
\newblock Approximate {B}ayesian inference for latent {G}aussian models by
  using integrated nested {L}aplace approximations.
\newblock {\em Journal of the royal statistical society: Series b (statistical
  methodology)\/}~{\em 71\/}(2), 319--392.
\newblock
  \href{https://doi.org/10.1111/j.1467-9868.2008.00700.x}{{10.1111/j.1467-9868.2008.00700.x}}.

\bibitem[\protect\citeauthoryear{Rue, Riebler, S{\o}rbye, Illian, Simpson, and
  Lindgren}{Rue et~al.}{2017}]{rue2017bayesian}
Rue, H., A.~Riebler, S.~H. S{\o}rbye, J.~B. Illian, D.~P. Simpson, and F.~K.
  Lindgren (2017).
\newblock Bayesian computing with{ INLA}: a review.
\newblock {\em Annual Review of Statistics and Its Application\/}~{\em 4},
  395--421.
\newblock
  \href{https://doi.org/10.1146/annurev-statistics-060116-054045}{{10.1146/annurev-statistics-060116-054045}}.

\bibitem[\protect\citeauthoryear{Rustand, {Van Niekerk}, Krainski, Rue, and
  {Proust-Lima}}{Rustand et~al.}{2022}]{art698}
Rustand, D., J.~{Van Niekerk}, E.~T. Krainski, H.~Rue, and C.~{Proust-Lima}
  (2022).
\newblock Fast and flexible inference approach for joint models of multivariate
  longitudinal and survival data using integrated nested {L}aplace
  approximations.
\newblock {\em (submitted)\/}~{\em xx\/}(xx), xx--xx.
\newblock \url{https://arxiv.org/abs/2203.06256}.

\bibitem[\protect\citeauthoryear{Saad}{Saad}{2003}]{saad2003iterative}
Saad, Y. (2003).
\newblock {\em Iterative methods for sparse linear systems}.
\newblock SIAM.
\newblock
  \href{https://doi.org/10.1137/1.9780898718003}{10.1137/1.9780898718003}.

\bibitem[\protect\citeauthoryear{Sanyal, Rochereau, Maesano, Com-Ruelle, and
  Annesi-Maesano}{Sanyal et~al.}{2018}]{sanyal2018}
Sanyal, S., T.~Rochereau, C.~N. Maesano, L.~Com-Ruelle, and I.~Annesi-Maesano
  (2018).
\newblock Long-term effect of outdoor air pollution on mortality and morbidity:
  a 12-year follow-up study for metropolitan france.
\newblock {\em International journal of environmental research and public
  health\/}~{\em 15\/}(11), 2487.
\newblock \url{https://doi.org/10.3390/ijerph15112487}.

\bibitem[\protect\citeauthoryear{Shaddick, Thomas, Amini, Broday, Cohen,
  Frostad, Green, Gumy, Liu, Martin, et~al.}{Shaddick
  et~al.}{2018}]{shaddick2018}
Shaddick, G., M.~L. Thomas, H.~Amini, D.~Broday, A.~Cohen, J.~Frostad,
  A.~Green, S.~Gumy, Y.~Liu, R.~V. Martin, et~al. (2018).
\newblock Data integration for the assessment of population exposure to ambient
  air pollution for global burden of disease assessment.
\newblock {\em Environmental science \& technology\/}~{\em 52\/}(16),
  9069--9078.
\newblock \url{https://doi.org/10.1021/acs.est.8b02864}.

\bibitem[\protect\citeauthoryear{Solomon}{Solomon}{2015}]{solomon2015numerical}
Solomon, J. (2015).
\newblock {\em Numerical algorithms: methods for computer vision, machine
  learning, and graphics}.
\newblock CRC press.
\newblock \url{https://doi.org/10.1201/b18657}.

\bibitem[\protect\citeauthoryear{Spencer, Yue, Bolin, Ryan, and Mejia}{Spencer
  et~al.}{2022}]{spencer2022spatial}
Spencer, D., Y.~R. Yue, D.~Bolin, S.~Ryan, and A.~F. Mejia (2022).
\newblock Spatial bayesian glm on the cortical surface produces reliable task
  activations in individuals and groups.
\newblock {\em NeuroImage\/}, 118908.
\newblock \url{https://doi.org/10.1016/j.neuroimage.2022.118908}.

\bibitem[\protect\citeauthoryear{Takahashi}{Takahashi}{1973}]{takahashi1973formation}
Takahashi, K. (1973).
\newblock Formation of sparse bus impedance matrix and its application to short
  circuit study.
\newblock In {\em Proc. PICA Conference, June, 1973}.

\bibitem[\protect\citeauthoryear{Toledo}{Toledo}{2003}]{toledo2003taucs}
Toledo, S. (2003).
\newblock Taucs: A library of sparse linear solvers.
\newblock \url{https://www.tau.ac.il/~stoledo/taucs/}.

\bibitem[\protect\citeauthoryear{Van~Merri{\"e}nboer, Breuleux, Bergeron, and
  Lamblin}{Van~Merri{\"e}nboer et~al.}{2018}]{van2018automatic}
Van~Merri{\"e}nboer, B., O.~Breuleux, A.~Bergeron, and P.~Lamblin (2018).
\newblock Automatic differentiation in {ML}: Where we are and where we should
  be going.
\newblock {\em Advances in neural information processing systems\/}~{\em 31}.
\newblock
  \url{https://proceedings.neurips.cc/paper/2018/file/770f8e448d07586afbf77bb59f698587-Paper.pdf}.

\bibitem[\protect\citeauthoryear{van Niekerk, Bakka, Rue, and Schenk}{van
  Niekerk et~al.}{2021}]{van2019new}
van Niekerk, J., H.~Bakka, H.~Rue, and O.~Schenk (2021).
\newblock New frontiers in {B}ayesian modeling using the {INLA} package in {R}.
\newblock {\em Journal of Statistical Software\/}~(1).
\newblock \url{https://arxiv.org/abs/1907.10426}.

\bibitem[\protect\citeauthoryear{Van~Niekerk, Krainski, Rustand, and
  Rue}{Van~Niekerk et~al.}{2022}]{van2022newinla}
Van~Niekerk, J., E.~Krainski, D.~Rustand, and H.~Rue (2022).
\newblock A new avenue for bayesian inference with inla.
\newblock {\em arXiv preprint\/}.

\bibitem[\protect\citeauthoryear{Verbosio, {De Coninck}, Kourounis, and
  Schenk}{Verbosio et~al.}{2017}]{VERBOSIO201799}
Verbosio, F., A.~{De Coninck}, D.~Kourounis, and O.~Schenk (2017).
\newblock Enhancing the scalability of selected inversion factorization
  algorithms in genomic prediction.
\newblock {\em Journal of Computational Science\/}~{\em 22}, 99--108.
\newblock \url{https://doi.org/10.1016/j.jocs.2017.08.013}.

\end{thebibliography}
%% if required, the content of .bbl file can be included here once bbl
%% is generated \input sn-article.bbl

% \input sn-sample-bib.tex%

\end{document}